\documentstyle[12pt,aaspp4]{article}
\tightenlines

\lefthead{Colbert \& Mushotzky}
\righthead{Black Holes in Nearby Galaxies}

\begin{document}

\title{The Nature of Accreting Black Holes in Nearby Galaxy Nuclei \\
}

\author{Edward J. M. Colbert \altaffilmark{1} and Richard F. Mushotzky}

\affil{Laboratory for High Energy Astrophysics, Code 662, NASA/GSFC,
Greenbelt, MD  20771}



\altaffiltext{1}{NAS/NRC Research Associate}

\def\amin{\arcmin}
\def\asec{\arcsec}
\catcode`\@=11
\def\gapprox{\mathrel{\mathpalette\@versim>}}
\def\lapprox{\mathrel{\mathpalette\@versim<}}
\def\@versim#1#2{\lower2.45pt\vbox{\baselineskip0pt\lineskip0.9pt
    \ialign{$\m@th#1\hfil##\hfil$\crcr#2\crcr\sim\crcr}}}
\catcode`\@=12
\def\etal{et al.\ }


\begin{abstract}
We have found compact X-ray sources in the center of 21 (54\%) of 39 nearby 
face-on spiral and elliptical galaxies with available ROSAT HRI data.  
ROSAT X-ray luminosities (0.2 $-$ 2.4 keV) of these compact X-ray
sources are $\sim$10$^{37}$$-$10$^{40}$ erg~s$^{-1}$ (with a mean of
3 $\times$ 10$^{39}$ erg~s$^{-1}$).  The 
mean displacement between
the location of the compact X-ray source and the
optical photometric center of the galaxy is $\sim$390 pc.
The fact that compact nuclear sources were found in nearly all (five of
six) galaxies with previous evidence for a black hole or an AGN
indicates that at least some of the X-ray sources are accreting
supermassive black holes.
ASCA spectra of six of the 21 galaxies show the presence of a hard
component with relatively steep ($\Gamma \approx$ 2.5) spectral slope.
A multicolor disk blackbody model fits the data from the spiral galaxies
well, suggesting that the X-ray object in these galaxies may be similar
to a Black Hole Candidate in its soft (high) state.  ASCA data from the
elliptical galaxies indicate that hot (kT $\approx$ 0.7 keV) gas 
dominates the emission. 
The fact that (for both spiral and elliptical galaxies)
the spectral slope is steeper than in normal type 1 AGNs and that
relatively low absorbing columns (N$_H \approx$ 10$^{21}$ cm$^{-2}$)
were found to the power-law component indicates that these objects are
somehow geometrically and/or physically different from AGNs in normal active 
galaxies.
The X-ray sources in the spiral and elliptical
galaxies may be black hole X-ray binaries,
low-luminosity AGNs, or possibly young X-ray luminous supernovae.  
Assuming the sources in the spiral galaxies are accreting black holes
in their soft state, we estimate black hole masses $\sim$10$^2$$-$10$^4$
M$_\odot$.
\end{abstract}




%

\section{Introduction}

An important unanswered question in extragalactic astronomy is whether
supermassive black holes (BHs) exist in the nuclei of most galaxies.
Since most galaxies are classified as ``normal,'' this question
translates into:  Are supermassive BHs present in the nuclei of most
normal galaxies?

It is commonly accepted that galaxies with active galactic nuclei (AGNs) 
have nuclear BHs with masses $\sim$10$^6$$-$10$^9$ M$_\odot$.
However, ``classical'' AGNs represent only a small fraction of all galaxies.
For example, Seyfert galaxies comprise
$\sim$1\% of all luminous spiral galaxies (Osterbrock 1989).
The geometrical model of AGNs is a supermassive
black hole immediately surrounded by a hot
accretion disk.  Further out ($\lapprox$ several pc), the hole and disk
are encircled by an optically thick molecular torus (cf., Antonucci
1993).  Optical spectra of AGNs show broad ($\sim$1000 km~s$^{-1}$ FWHM)
and ``narrow'' (several 100 km~s$^{-1}$ FWHM) emission-line features
from, respectively, the so-called broad-line region (BLR) located
within $\sim$0.1 pc, and the narrow-line region (NLR), which lies out to
radii of several 100 pc.  
Convincing evidence for nuclear BHs in active galaxies comes from rotation 
curves of H$_2$O megamaser sources in the LINER galaxy NGC~4258 (Miyoshi
et al. 1995) and rotation curves of a small, ionized gas disk  
in the nearby radio galaxy M87 (Ford et al. 1994; Harms et al. 1994).

There is considerably less known about the existence of supermassive BHs
in normal galaxy nuclei.
Dynamical searches for BHs in normal galaxies reveal central
dark objects with masses $\sim$10$^6$$-$10$^{9.5}$ M$_\odot$ in many
nearby galaxies (cf. Kormendy \& Richstone 1995), so it is feasible that
many normal galaxies do contain nuclear supermassive BHs.  However,
normal galaxy nuclei do not show optical evidence for BLRs and NLRs and 
their optical/ultraviolet continua do not typically show the signature of the 
``big blue bump,'' which is thought to be emission from the accretion
disk in AGNs.  One possible model of galaxies purports that 
supermassive BHs
are present in all galaxies, but are ``active'' in some and ``inactive''
in the others (e.g. Cavaliere \& Padovani 1988).  In this scenario, normal
galaxies could be galaxies in which the BH/accretion disk
is relatively ``inactive.''

Why might supermassive BHs in normal galaxies be less 
luminous than AGNs in active galaxies?  There are at least two
possibilities. 
First, the mass of the BH may be significantly smaller 
(M$_{BH} \ll$ 10$^6$ M$_\odot$).  Second, the accretion rate may be much less
(than in Seyfert galaxy AGNs, for example).  
It has been proposed by Narayan \& Yi (1994, 1995a, 1995b) that
accreting flows in BHs with low accretion rates are advection dominated
(i.e., very little energy is radiated away during the accretion
process).  In this model for advection-dominated accretion flows
(ADAFs), matter does not flow inwards in the form of a standard thin
accretion disk, the flow pattern is closer to being spherical.
An ADAF model fits the observed properties of the low-luminosity AGN NGC~4258,
although a thin disk plus corona model is also allowed
Lasota et al. 1996; Herrnstein et al. 1998; Gammie, Narayan \& Blandford
1999).
The X-ray emission from cores of elliptical galaxies is also well
modelled by an ADAF and is consistent with central black hole masses of
$\sim$10$^9$ M$\odot$ (Reynolds et al. 1997; Di~Matteo \& Fabian 1997
and references therein), as is observed for M87 (Ford et al. 1994; Harms
et al. 1994).

X-ray emission from accreting BHs with standard thin accretion disks has a
characteristic power-law spectrum, thought to be produced by
Compton scattering of ultraviolet photons from the accretion disk (cf.
Svensson 1996).  
ADAF accretion flows produce soft X-ray emission by Comptonization of
synchrotron photons and hard X-rays 
via bremsstrahlung (Mahadevan 1997).  Whether 
accreting BHs in normal galaxy nuclei are surrounded by a thin accretion
disk or are advection-dominated, they should be associated with compact
X-ray emission from regions immediately surrounding the BH, so 
compact X-ray sources are good indicators of nuclear BHs in galaxy nuclei.
Only a few different types of point sources in external galaxies emit at
high enough fluxes so that present day 
X-ray observations can detect them: X-ray
bright supernova remnants (cf. Schlegel 1995), luminous X-ray binaries,
and luminous galactic nuclei.  At other wavelengths (e.g., optical),
there are much more possible sources of emission besides compact
galactic nuclei.  Therefore, X-ray observations are unique in their
ability to identify accreting BHs in galactic nuclei.

As an example, consider the X-ray source in 
the nuclear region of the nearby normal galaxy NGC~1313.
A bright X-ray source in the nuclear region of this galaxy is
located $\sim$1 kpc from the optical nucleus, has 
L$_X(0.2-2.4~{\rm keV}) \sim$ 10$^{40}$
erg~s$^{-1}$, implying a BH mass of $\sim$10$^3$$-$10$^4$ M$_\odot$
for
an Eddington ratio of 10$^{-2}$$-$10$^{-1}$ (Colbert et al. 1995).
There is no evidence for an AGN
at wavelengths other than X-ray, yet
the implied mass is much larger than those of typical BH X-ray binaries
(XRBs)
($\sim$5$-$10 M$_\odot$, Tanaka 1989).  

We seek to understand the
nature of these near-nuclear compact X-ray sources.  Do they represent a
class of low-luminosity AGN perhaps similar to Seyfert nuclei, e.g. AGNs with
advection-dominated accretion flows (ADAFs)?  
Or are they BH XRBs with BH masses of $\sim$10$-$10$^4$ M$_\odot$ ?
Or are they a new class of object with properties yet to be discovered?


In the present paper, we describe the X-ray properties of compact X-ray
sources in the nuclei of a sample of nearby normal galaxies, with the
hope of determining the nature of the X-ray sources.  In section 2, we
describe the selection process for the sample, and in section 3 we
describe the observations and data reduction technique.  The results
from the ROSAT HRI survey (of 39 galaxies) 
are presented in section 4 and the results from ASCA observations of six
of those galaxies are presented in section 5.  Possible scenarios for the 
origin of the compact X-ray sources are discussed in section 6.  A summary 
of the results is given in section 7.

\section{Sample Selection}

Our goal was to select a sample of galaxies in which compact
X-ray sources could be associated with accreting BHs as unambiguously as
possible.  By using data from the ROSAT HRI instrument, we choose the
the best angular resolution X-ray data available for such a study.
All galaxies with recessional velocities smaller than 1000 km~s$^{-1}$
were selected from an electronic version (vers. 3.9) of the 
{\it Third Reference Catalog of Bright Galaxies} 
(de~Vaucouleurs \etal 1991;RC3).  
We omitted galaxies
in which the disks were edge-on, since the additional large absorbing
columns significantly lower 
the count rates of any potential X-ray sources and might produce
low-quality spectra.  More specifically, we required that
the logarithm of the axial ratio be smaller than 0.4 (as taken from
RC3).  Starburst galaxies produce X-ray emitting galactic winds which
extend out of the nuclear region and the nuclear starbursts themselves
are sources of X-ray emission (often with very complex structure), 
so we chose to omit starburst galaxies
from our sample.  We deselected starburst galaxies by requiring L$_{FIR}$
(calculated from IRAS fluxes using the method described in Fullmer \&
Lonsdale 1989) to be less than the blue B luminosity L$_B$.  We then 
required that there be sufficiently good (exposure time $\gapprox$ 3 ks) 
ROSAT HRI data available in the
public archives, so that luminosities $\gapprox$ 10$^{39}$ erg~s$^{-1}$
could be sampled. 
The Seyfert galaxy NGC~4151 was omitted from the sample since it is
already known to have a bright 
X-ray emitting AGN.  The Local Group galaxies SMC and LMC
were also omitted from the sample since they project too large an area on
the sky to make ROSAT imaging feasible without mosaicing.
These selection 
criteria produced a list of 39 nearby galaxies (see Table \ref{tabros39}).

\placetable{tabros39}

\section{Observations and Data Reduction}

\subsection{ROSAT HRI data}

All available HRI data for the selected 39 galaxies were retrieved from the 
US ROSAT archives at NASA/GSFC.  
A list of the observational data used is given in Table \ref{tabrosobs}, 
along with the exposure times.
We used the IRAF/PROS software package
to produce event list files, which were then used for imaging.  All
imaging data were first blocked by a factor of two to give 1" pixels,
and were then smoothed with a Gaussian profile with $\sigma =$ 3\asec.
Contour maps of the smoothed images were overlayed on digitized B-band
survey plate images (from the STScI digitized sky survey) in order to
visualize the relative location of the compact X-ray sources with
respect to the galaxy nucleus.  We defined an X-ray source to be in the
``center'' of the galaxy (i.e., a nuclear source) if it was located
within 2 arcminutes of the optical photometric center.

\placetable{tabrosobs}

Count rates were calculated from the unsmoothed HRI images using circular 
source regions of radius 15", centered on the peak pixel.  Background counts 
were taken from annular
regions surrounding the source.  It was necessary in a few cases to
restrict the azimuthal angle of the background region so that there
would not be any contamination from nearby sources.

For the galaxies M33, M81, NGC~4374, NGC~4406 and NGC~4552, diffuse X-ray
emission was present in the nuclear region, i.e., in the circular source
regions.  
This was determined by trying to fit radial profiles for all of the galaxies
with a Gaussian plus a constant and examining the residual X-ray
emission.  All but these five galaxies were consistent with the HRI
point-spread function.
Since the background region also samples the diffuse emission,
the net count rate was partially corrected when we subtracted the normalized 
background counts from the source counts in the source region.
The diffuse emission tends to fill the source regions
more than it fills the background region, so there is still some
contaminated emission accounted for in the net count rate. 




For the galaxies with extended diffuse emission, we estimated the count
rate from the pointlike component of the X-ray emission by modelling the
radial profile from the HRI image.  Radial profiles were extracted using
the XIMAGE software and models were fit to the data using the QDP
plotting package.  We fit a model consisting of a Gaussian component for
the point-spread function of the compact X-ray source, a King model for
the extended X-ray emission, and a constant component for the background
X-ray emission.  We then integrated the Gaussian and King components out
to a radius of 15\asec\ (the size of the circle used for extracting
source counts from the HRI image) and these integrals were used to
determine the fraction of the total source counts which were from the 
(Gaussian) point source.  We found fractions of 0.373, 0.611, 0.151,
0.328 and 0.171 for M33, M81, NGC~4374, NGC~4406, and NGC~4552,
respectively.  The net count rates for these five galaxies were modified
using these correction factors and fluxes and luminosities in this paper
correspond to these corrected count rates.

The ROSAT HRI data for the individual galaxies was also checked for
variability.  XSELECT was used to extract light curves from the raw data
using the same source regions described above.  XRONOS was used to plot
the light curves.  Most of the light curves showed less than 50\%
variability, althought two galaxies showed variability on scales of
50\% or more (Ho~II displayed $\sim$50\% variability on $\sim$0.2 day
time scales and NGC~5204 displayed $\sim$100\% variability on $\sim$0.3
day time scales).

\subsection{ASCA data}

Data for eight observations (of six galaxies) in our sample were
retrieved from the ASCA archives in order to facilitate our
understanding of the nature of the compact X-ray sources.  The six
galaxies were selected because (1) they have compact X-ray sources in
the HRI images, (2) no adjacent X-ray sources were present in the
ASCA fields (so that the ASCA spectra include point-source and possible
diffuse emission only from the galaxy
under study), and (3) public archival ASCA data 
were available.  A list of the ASCA observations is
given in Table \ref{tabascaobs}.

\placetable{tabascaobs}

The raw (``unscreened'') ASCA data were screened using the XSELECT
software program (part of the FTOOLS X-ray analysis software package)
using the following restrictions.  Only BRIGHT mode data were used with
high, medium and low bit rates.  The maximum angular deviation from the
source was set at 0.01$^\circ$, the angle from bright Earth was
restricted to be greater than 20$^\circ$, the minimum elevation angle was
set at 5$^\circ$ for the GIS data and 10$^\circ$ for the SIS data, the minimum
cutoff rigidity (COR) was set to be 6 GeV/c, the Radiation Belt Monitor
was restricted to be below 200, and the SIS PIXL rejection threshold was
set at 100, 75 and 50 for 1, 2 and 4 CCD modes, respectively.  

Source counts were extracted using circular or polygonal regions with
typical diameters of $\sim$6$-$8' (SIS) and $\sim$7$-$12' (GIS).  For
both observations of NGC~1313, the central (nearest to the nucleus)
X-ray source fell near the corner of the SIS CCD chip and polygonal
regions with diameters $\sim$4$-$5' were used to extract source counts.
Consequently, the SIS fluxes for NGC~1313 include a fraction (estimated
$\gapprox$ 60\% from the ASCA point spread function)
of the total emission from the source.  Background
counts were taken from source-free polygonal regions surrounding the
source regions.  For the SIS spectra, background counts were taken from
a region of the same CCD in which the source was located.  Extended
X-ray emission was noticed in the GIS image of NGC~4406, thus the source
counts for that galaxy include some of the emission from this extended
X-ray source.

\section {Results from ROSAT HRI Observations}

\subsection{Notes on Individual Galaxies}

Here we list notes for each of the 39 galaxies in the sample.  The
galaxies are organized in three different groups: dwarf galaxies,
spiral and irregular galaxies, and elliptical and lenticular galaxies.

\subsubsection{Dwarf Galaxies}

{\bf NGC~205 (M110)} -- This is a normal dwarf galaxy and is
one of the two elliptical companions to M31.  There
is no previous evidence for a central black hole and we do not detect
a compact X-ray source.

{\bf NGC~221 (M32)} -- 
M32 is the other normal dwarf elliptical galaxy that is a companion to
M31.  Eskridge, White \& Davis (1996) analyzed the ROSAT PSPC data for M32 and
concluded that the emission from the central source could be due to
either low-mass X-ray binaries or accretion onto a massive central black
hole.  Loewenstein et al. (1998) argue that the central source is most
likely a single X-ray binary rather than an AGN, based on the presence
of an offset of the compact X-ray source (offset $=$ 33.6 pc, from our 
HRI image) with
respect to the nucleus and extremely large intensity variations (from
ASCA observations).  Previous evidence for a central black hole of mass
10$^{6.5}$ M$_\odot$ comes from dynamical evidence (e.g., Bender et al. 1996).

{\bf NGC~1705} --  Lamb et al. (1985) suggest that this dwarf galaxy is
in a starburst or post-starburst phase.  NED lists its classification as
an H~II region galaxy.  We did not detect any compact X-ray
sources in the nuclear region of NGC~1705.

{\bf Ho~II} --  This dwarf Irr galaxy is a companion to M81.  We found
a bright compact X-ray source approximately 2 arcminutes to the east of the
galaxy center.

{\bf ESO~495-G21 (He 2-10)} --  ESO~495-G21 was originally discovered as
a planetary nebula (He~2-10) and is now known as a dwarf emission-line 
galaxy.  We find a compact X-ray source coincident with the nucleus.

{\bf UGC~5086} --  We did not detect any
X-ray emission from this dwarf Irr galaxy.

{\bf Ho~IX} -- This is a dwarf Irr galaxy and is a companion to M81.  We
did not find any X-ray emission from this galaxy.

{\bf Leo~A} --  This dwarf Irr galaxy is a possible member of the Local
Group.  We did not find any X-ray emission from this galaxy.

{\bf IC~3303} --  We found no X-ray emission from this dwarf spiral
galaxy.

{\bf NGC~4627} --  NGC~4627 is a dwarf elliptical galaxy and is a nearby
companion to the edge-on spiral galaxy NGC~4631.  We did not find any
X-ray emission from NGC~4627.

{\bf NGC~5204} --  This dwarf galaxy is labeled as having an H~II region
nucleus by HFS95.  We find a compact X-ray source offset 17.3" (403 pc)
from the nucleus.


\subsubsection{Spiral and Irregular Galaxies}

{\bf NGC~300} -- 
This is a normal spiral galaxy similar to M33 in appearance, located in
the Sculptor group.  We did not find any compact X-ray sources within
1' of the nucleus of NGC~300.

{\bf NGC~598 (M33)} --
M33 is a normal spiral galaxy in the Local Group.  Of all of the
galaxies in the Local Group, it has the most luminous X-ray nucleus.
Recently Dubus et al. (1997) have found evidence for 106-day periodic 
variability of the nuclear X-ray source and argue that the source is a
black hole X-ray binary with a $\sim$10 M$_\odot$ primary (for an
Eddington ratio of unity).  The ASCA spectra of the central X-ray source
have been analyzed by Takano et al. (1994).  We discuss ASCA
observations of this X-ray source in section 5.

{\bf NGC~1313} --
NGC~1313 is a nearby normal spiral galaxy with many H~II regions.
Colbert et al. (1995) found that the central bright X-ray source was
offset $\sim$1 kpc from the photometric center, from analysis of ROSAT
PSPC data.  They argued that the X-ray source was an accretion-powered
black hole of mass $\sim$10$^3$$-$10$^4$ M$_\odot$ (for an Eddington
ratio $\sim$ 10$^{-2}$$-$10$^{-1}$), but unlike an AGN since there is no
evidence for a BLR, NLR or accretion disk.  ASCA observations of the
central X-ray source in this galaxy are discussed in section 5.
A previous analysis of one of the ASCA observations was published in
Petre et al. (1994).

{\bf AM~0337-353} --  This galaxy is a normal spiral galaxy, also
located in the Fornax cluster.  We did not detect any compact X-ray
sources in the nuclear region of AM~0337-353.

{\bf IC~342} --  IC~342 is a large, face-on spiral galaxy with a normal
galaxy nucleus.  We found several compact X-ray sources in IC~342, one
of which was coincident with the galaxy nucleus.  ASCA observations of
IC~342 are unable to isolate the central X-ray source seen in the ROSAT
HRI image from other surrounding X-ray sources in the galaxy (Okada et
al. 1994).  However, Okada et al. (1998) suggest that another X-ray 
source (not the most central one) in the central regions of IC~342 may 
be an accreting BH of mass $\sim$100 M$_\odot$.

{\bf NGC~2403} --  NGC~2403 is a nearby spiral galaxy with similar
global properties to M33 (Thornley 1996).  The galaxy has many H~II
regions, but the nucleus seems quiescent (Beckman et al. 1987).  A
compact X-ray source we found was near (offset $=$ 17.1"), but not coincident 
with the galaxy nucleus.

{\bf NGC~3031 (M81)} --
This galaxy is described as having a LINER nuclear spectrum (Heckman
1980); some nuclei classified as LINERs are low-luminosity AGNs (e.g., 
Ho, Filippenko, \& Sargent 1994, hereafter HFS95).
Petre et al. (1993) describe its X-ray spectrum as being similar to that
of other LINERs with broad H$\alpha$ emission, but argue against the
presence of an accretion disk.
Ishisaki et al. (1996) argue for the presence of a low-luminosity AGN
in M81, based on the results from eight ASCA observations.

{\bf UGCA~276} --  We found no X-ray emission from this Irr galaxy.

{\bf UGC~7356} --  We found no X-ray emission from this Irr galaxy.

{\bf NGC~4286} -- No X-ray emission was found from this nearby spiral
galaxy.

{\bf NGC~4449} --  This is a Magellanic Irr galaxy with many H~II
regions and is undergoing active star formation (Bothun 1986).  We found
multiple compact X-ray sources in this galaxy, none of which is
coincident with the galaxy center.  The nearest X-ray source to the
nucleus is located 39.3" (572 pc) from the nucleus.

{\bf NGC~4485} --  This Irr galaxy is tidally interacting with NGC~4490.
HFS95 classify it as having an H~II region nucleus.  We detect very
faint emission that is offset 20.5" (924 pc) from the galaxy center.

{\bf NGC~4569 (M90)} --  This spiral galaxy is in the Virgo cluster.
HFS95 label it as having a transition (LINER/H~II) type nucleus.
We find a compact X-ray source coincident with the nucleus.

{\bf NGC~5408} -- NGC~5408 is an Irr galaxy.  We detected a compact
X-ray source offset 49.3" (1.2 kpc) from the galaxy center.  We discuss
the ASCA spectra of this X-ray source in section 5.

{\bf NGC~6822} -- NGC~6822 is an Irr galaxy in the Local Group.  We
detected a compact X-ray source offset 28.4" (96.4 pc) from the galaxy
center.

{\bf NGC~6946} --  This is a face-on spiral galaxy which has an H~II
region nucleus (HFS95).  We find several compact X-ray sources
associated with NGC~6946, the closest one to the nucleus being 
coincident (offset $<$ 10") with the galaxy nucleus.

{\bf NGC~7320} --  This is the brightest spiral galaxy in Stephan's
quintet.  We did not find any X-ray emission from NGC~7320 (see also
Pietsch et al. 1997).


\subsubsection{Elliptical and Lenticular Galaxies}

{\bf NGC~1396} --  This galaxy is a normal lenticular/elliptical galaxy
located in the Fornax cluster.  We did not detect any compact X-ray sources
in the nuclear region of NGC~1396.

{\bf NGC~3990} -- NGC~3990 is a lenticular galaxy.  We found no X-ray
emission from this galaxy.

{\bf NGC~4138} -- This spiral galaxy has a nucleus classified as ``S1.9''
by HFS95.  However, we found no X-ray emission from NGC~4138.

{\bf NGC~4150} -- NGC~4150 is a nearby, small lenticular galaxy.  We
found a compact  X-ray source offset by 16.3" (767 pc) from the galaxy 
nucleus.

{\bf NGC~4190} -- NGC~4190 is another nearby, small lenticular galaxy.
The compact X-ray source we found is located 7.4" from the optical
galaxy center.

{\bf NGC~4278} --  This is a nearby elliptical galaxy with large-scale
radio emission (Wrobel \& Heeschen 1984).  NED lists its classification
as a Seyfert~1 galaxy and HFS95 classify its nucleus as having a LINER
spectrum.  We find a compact X-ray source coincident with the galaxy
nucleus.

{\bf NGC~4374 (M84)} -- M84 is a classical example of a radio galaxy
(cf. Bridle \& Perley 1984) and is hosted by a giant elliptical galaxy.
We find an X-ray source coincident with the nucleus and also find
diffuse X-ray emission surrounding the nuclear X-ray source.
The ASCA spectra of this X-ray source are discussed in section 5.

{\bf NGC~4387} --  This elliptical galaxy did not have any detectable
X-ray emission.

{\bf NGC~4406 (M86)} --
M86 is a bright elliptical galaxy in the Virgo cluster.  We find a weak
nuclear X-ray source and also detect the plume of diffuse X-ray emission
extending toward the northwest (Forman, Jones \& Tucker 1985).
We discuss the ASCA spectra of this galaxy in section 5.

{\bf IC~3540} -- We did not find any X-ray emission from this S0
galaxy.

{\bf NGC~4552 (M89)} --  This elliptical galaxy is also quite a bright
radio source (Wrobel \& Heeschen 1984).  HFS95 label it as a transition
object.  We find a central compact X-ray source coincident with the 
nucleus, surrounded by diffuse X-ray emission.
The ASCA spectra of this source are discussed in section 5.


\subsection{Overview of Results from HRI survey}

Of the 39 galaxies for which HRI imaging data were available, we found 
(detection $>$ 3$\sigma$)
compact X-ray sources within $\sim$2\amin\ of the nucleus of 21
galaxies.  Lira, Lawrence \& Johnson (1998) find a similar result.
Here we have used the optical position listed in RC3 (the 
photometric center) as the location of the nucleus.  Estimated uncertainties
in the optical positions are given in RC3 as 0.$^s$1 and 1".
Upper limits (3$\sigma$) for the galaxies in which no sources were found
are listed in Table \ref{tabrosundetect}.  The X-ray
sources in the 21 galaxies in which compact sources were detected 
are listed in Table \ref{tabrosdetect}.

\placetable{tabrosundetect}
\placetable{tabrosdetect}

In Table \ref{tabrosdetect}, we 
list the X-ray position of the compact X-ray source, the
offset of the X-ray position from the photometric center, the HRI net
count rate, and Galactic absorption column in the direction of the X-ray
source.  Also listed are (unabsorbed, e.g., corrected for Galactic
absorption) 0.2$-$2.4 keV X-ray fluxes and luminosities of
the X-ray source, assuming a power-law (photon index $\Gamma =$ 1.7) spectrum 
and absorption from the Galactic hydrogen column.
Note that the count rates, X-ray fluxes and luminosities for the
galaxies with extended emission (M33, M81, NGC~4374, NGC~4406, and
NGC~4552) have been corrected so that they reflect only the contribution
from the point-like X-ray component (see section 3.1).

\subsection{Host Galaxy Type}

The host galaxies of the compact X-ray sources range from dwarf
irregular to elliptical to spiral.  We show histograms of the host
galaxy type (from RC3) in Figure \ref{fig1}.
The upper panel corresponds to galaxies in our sample with detected
compact X-ray sources, while the lower panel corresponds to all galaxies
in our sample.
Elliptical galaxies have galaxy
types -6, -5 or -4, lenticular galaxies -3, -2, or -1, spirals
0 {...} 9, and irregulars 10, 11 and 90.  One galaxy (ESO~495-G21, T =
90.0) has be omitted from the histogram for clarity.  Although there
seems to be
a slight preference for the X-ray sources to be found either in late-type
spirals and irregular galaxies, or elliptical galaxies, as can be seen
in the lower plot, our sample is abnormally deficient in early-type
spiral galaxies.  Histograms for complete samples from RC3 are flat
across galaxy T-type (e.g., Buta et al. 1994).

\placefigure{fig1}

\subsection{X-ray Luminosities}

A histogram of the 0.2$-$2.4 keV luminosities is shown in Figure \ref{fig2}.
Luminosities of the compact sources ranged from 8 $\times$10$^{36}$
erg~s$^{-1}$ to $\sim$10$^{40}$ erg~s$^{-1}$, with a mean of 
3 $\times$ 10$^{39}$ erg~s$^{-1}$.
For reference, Seyfert galaxies typically
have L$_X \sim$ 10$^{42}$$-$10$^{44}$ erg~s$^{-1}$ and the Eddington
luminosity of a 1.4 M$_\odot$ accreting neutron star is 10$^{38.3}$
erg~s$^{-1}$.
So, on average, the X-ray sources are at least one order of magnitude
more luminous than the most luminous (i.e., Eddington ratio of unity) 
neutron star X-ray binaries, but are several orders of magnitude less
luminous than typical AGN in Seyfert nuclei.  

Also shown in Figure 2 are the galaxy types of the host galaxies.  Note
that the most luminous X-ray sources (L$_X \sim$ 10$^{40}$
erg~s$^{-1}$) tend to be found in elliptical hosts.

\placefigure{fig2}

\subsection{Separation from Galaxy Center}

As can be noted from the entries in Table \ref{tabrosdetect}, 
the offsets of the X-ray
sources from the galaxy nucleus (photometric center) is significant in
many of the sources.  In Figure \ref{fig3}, we show a histogram of the angular
separation between the X-ray source and photometric center.
An interesting result of our study is that $\sim$40\% (9 of 21)
of the compact sources were located $\gapprox$ 10" 
($\gapprox$ 0.3 kpc at the mean distance of our sample)
from the
photometric optical center (as listed in RC3).  For comparison, most
observed objects have ROSAT X-ray positional errors of $\lapprox$10"
(Briel et al. 1996).

\placefigure{fig3}

In Figure 4, we show a plot of L$_X$ vs. the angular separation between
the X-ray source and the photometric center.  It is noteworthy that the
mean luminosity of sources with small ($\lapprox$200$-$400 pc)
separation is similar to the mean luminosity of sources with large
separation ($\gapprox$200$-$400 pc; see also section 6.2.2).

\placefigure{fig4}

\section{Results from ASCA Spectral Fitting}

As noted previously, ASCA spectra were extracted for six galaxies in the
sample (see Table \ref{tabascaobs}).  
We attempted to fit several spectral models to
the data using all four ASCA instruments (SIS0, SIS1, GIS2 and GIS3)
simultaneously.  We allowed the relative normalization of SIS1, GIS2 and
GIS3 (with respect to SIS0) to vary while fitting each model.
We ignored data from spectral bins marked ``bad'' by the XSELECT program
as well as data from energy bins below 0.4 keV and above 10.0 keV.

We first tried fitting the data with two single component models: (1)
a Raymond-Smith (RS) thermal plasma (Raymond \& Smith 1977), as would be
expected if the source were hot gas, and (2) a power-law (PL) model, as
might be expected from an AGN or other object accreting matter onto a
BH via an accretion disk.  The results from these fits are listed in
Table \ref{tabascasingle}.  
We list the spiral and elliptical galaxies separately since
elliptical galaxies often show evidence for a strong component of hot
gas (e.g. Forman et al. 1979), while spiral galaxies do not 
necessarily show such evidence.

\placetable{tabascasingle}

A cursory look at the results in Table \ref{tabascasingle} shows 
that the power-law model
is a better fit, in general, to the X-ray sources in the spiral
galaxies, whereas the Raymond-Smith plasma model is a better fit for the
elliptical galaxy sources.  The power law fit for the elliptical
galaxies is so poor for NGC~4374 and NGC~4406 that the reduced $\chi^2$
values are greater than 2.0, which is too poor to be able to be able to
use $\chi^2$ fitting to deduce standard errors for the parameters.
It is immediately noticeable that the first
observation of NGC~1313 gives a much different fit than for the second
observation of NGC~1313.  Most of the spiral galaxy sources are fit by a
steep power law with $\Gamma \approx$ 2.3 $-$ 2.8 or a thermal plasma
with kT $=$ 2.0 $-$ 3.4 keV.  However, in the first observation of
NGC~1313, the X-ray  source is best fit
with a significantly shallower power-law spectrum ($\Gamma =$ 1.7)
 or a hotter Raymond-Smith plasma with kT $=$ 6.7 keV.

The single-component fits to the elliptical galaxy sources are much
poorer than those to the spiral galaxy sources, with $\chi^2_\nu >$ 1.5
for five out of six fits.  On the other hand, the single-component fits
to the spiral galaxy sources are acceptable in nearly all cases, with 
$\chi^2_\nu >$ 1.2 for only two out of ten fits.

We also tried modelling the data with multiple-component fits, for
example with a RS+PL model (as appropriate for a BH embedded in diffuse
gas, e.g., Ptak et al. 1998), and a disk blackbody (DBB, Mitsuda et al.
1988) plus power-law model (DBB+PL, as is seen in Galactic BH
candidates).  For the RS+PL model, the RS 
component was assumed to only be absorbed by gas in our Galaxy (for
Galactic columns, see Table \ref{tabrosdetect}), 
while absorption to the PL component
was allowed to be larger.  For the DBB+PL fit, the absorption to 
both components was assumed to be the same and was allowed to be
larger than the Galactic value.  The results from these fits are
shown in Table 7. 

\placetable{7} 

We also list in Table 7 
the values of the F statistic, comparing the
single-component fits to the two-component fits.  In nearly all cases,
the two-component fits are preferred with very low probabilities of 
exceeding the F values by chance (significance of the extra terms
$\gapprox$ 97 $-$ 99.9\%).  An exception is M33, in which the single
component Raymond-Smith thermal plasma
model is nearly equally as good a fit as the DBB+PL model.
It should be noted here that the presence of a second (hard)
component in X-ray
spectra of elliptical galaxies has previously been discovered in ASCA 
spectra by Matsushita et al. (1994).

It is worthwhile to note that the elliptical galaxies are much better
fit by a RS+PL model compared to a DBB+PL model, indicative of
a hot gas component which is known to be common in these galaxies.
Furthermore, the DBB fits to the elliptical galaxies gave very 
large values for the inner
radius of the accretion disk (r$_{in}(\cos\theta)^{{1}\over{2}}$) and imply
considerably steeper power-law indices (compared to those from the RS+PL
fits) and very low values of kT$_{in}$
(the temperature at the inner radius of the accretion disk).  This
suggests that the DBB+PL model is incorrect for the elliptical galaxies
and another component (or a different component) is needed to fit the
spectra.  For these
reasons, we tried an additional three-component model consisting of a
Raymond-Smith thermal plasma, a DBB, and a power-law model.  The results
of these fits are given in Table 8. 

\placetable{8} 

The values of the F-statistic for comparison with the two-component fits
are also listed in Table 8. 
As can be seen from these values, the
RS+DBB+PL fits are 
significantly better
fits than the DBB+PL fits, but
are not significantly preferred over the RS+PL models.  Thus, we
conclude that a disk blackbody component is not required by the spectral
data and the RS+PL fits are the best fit models for the elliptical
galaxies in our sample.

We should also note that for two (NGC~4406 and NGC~4552) of the three
elliptical galaxies, the GIS spectra have higher normalization values
than the SIS spectra (by mean values of 1.9 and 1.5, respectively).
Since the GIS extraction regions are larger than those for the SIS, this
implies that there is an extended component of X-ray emission, most
likely a hot thermal plasma which we have modelled with a Raymond-Smith
component.
The ROSAT HRI images of these galaxies confirm that extended X-ray
emission is present surrounding the central X-ray source (section 4.1).

The power-law slopes for the elliptical galaxy
RS+PL models are considerably steeper than
found for ASCA analysis of type 1 AGN, in which $<\Gamma> =$ 1.91 $\pm$
0.07 (Nandra et al. 1997).
The mean value and standard deviation for our three elliptical galaxies
is 2.55 $\pm$ 0.76.  This may be indicative of a different kind of
emission process, for example both Narrow-Line Seyfert 1 galaxies and
BH XRBs in their soft state have comparably steep power-law slopes.

We next turn to the three spiral galaxies in our ASCA sample.  M33 is
best fit by a DBB+PL model (as previously noted by Takano et al. 1994),
although a RS+PL model is also allowed statistically ($\chi^2_\nu$
values of 0.98 and 1.00 for the two separate observations).  NGC~1313
and NGC~5408 are nearly equally well fit by either a RS+PL model or a
DBB+PL model (the RS+PL model yields a slightly better fit).

Again, it is noteworthy that the power-law slopes for the spiral
galaxies are also comparatively steep when compared with ASCA spectra of
Seyfert 1 galaxies.  

We next discuss the results of the spiral galaxy spectra in terms of
emission expected from an BH XRB.  In the hard state, BH XRBs typically have
hard spectra with power-law slopes $\Gamma \approx$ 1.3 $-$ 1.9, whereas
in the soft state, the power-law slope steepens to $\approx$ 2.5 
(e.g., Ebisawa, Titarchuk, \& Chakrabarti 1996) and a 
soft component appears in the X-ray spectrum (usually modelled as
a multicolor disk blackbody component).  The first observation of
NGC~1313 exhibits a slope more consistent with a BH XRB in its hard
state, while in the second observation the slope steepens to $>$2.5,
more consistent with a soft state.

For the spiral galaxy X-ray sources, 
the values of the inner radius of the accretion disk (as predicted by
the multicolor disk blackbody model) are given in Table \ref{tabrincostheta}.
For M33, 
r$_{in}(\cos\theta)^{{1}\over{2}}$ ($\approx$ 40 km) 
is more consistent with values
for BH candidates ($\approx$ 20 $-$ 30 km) than for low-mass X-ray
binaries (LMXBs)
($\approx$ 5 $-$ 10 km; e.g., Yaqoob, Ebisawa, \& Mitsuda 1993).  
The second observation of NGC~1313
(for which a steep spectral slope suggests a soft state) also yields a
value of r$_{in}(\cos\theta)^{{1}\over{2}}$ ($\approx$ 30 km), consistent 
with a BH candidate.
However, for NGC~5408, a significantly larger inner
radius (r$_{in}(\cos\theta)^{{1}\over{2}}$ $\approx$ 18000 km)
is predicted.  For comparison, the Schwarzschild radius 2GM/c$^2$
of a 10$^3$ M$_\odot$ BH is equal to 2950 km).  Although we cannot show
that the objects in these galaxies exhibit XRB-like emission, we can say
that if the X-ray emission is disk blackbody emission from an XRB-like
accretion disk in its soft state, the inner radii imply 
that the objects are BH candidates.

\placetable{tabrincostheta}

To further investigate the hypothesis that the emission is from XRB-like
objects in their soft state, we fitted the spectra from the spiral galaxies
with the bulk-motion (BM) model promoted by Titarchuk and collaborators
(e.g., see Shrader \& Titarchuk 1998).  
As in the DBB+PL model, in this model, the soft X-ray photons are assumed
to come directly from the thin accretion disk, whereas the hard photons
come from upscattering of the soft photons by bulk motion and from
downscattering of harder photons by Comptonization.
Results from these fits are
shown in Table \ref{tabascabm}.

\placetable{tabascabm}

In general, the fits were acceptable, although the reduced $\chi^2$
values were typically higher than those for 
the RS+PL and DBB+PL fits.  Again, the
power-law slopes predicted by the BM model are quite steep ($\sim$ 2.5,
except for NGC~1313 observation 1, which seems to be in a hard state).
Such steep slopes are consistent with a BH XRB in its soft state.

The normalization parameter from fitting the Bulk Motion model
allows us to directly estimate
the masses of the central BH.  We list those predicted masses in 
Table \ref{tabrincostheta}.  
Note that for M33 and NGC~1313\#2, these calculations imply 
BH candidates with masses $\gapprox$ 10$^2$ M$_\odot$.
For NGC~5408,
the predicted mass of the BH is $\sim$10$^4$ M$_\odot$.  We find
that the best fit value of r$_{in}(\cos\theta)^{{1}\over{2}}$ 
(from the DBB+PL models; Table \ref{tabrincostheta}) is less
than the corresponding r$_g$ for with the mass predicted by the
BM models. 
We discuss this further in section 6.5.


We note that for the spiral galaxies, the DBB+PL model requires an
additional (above that for our Galaxy) absorbing column of 
N$_H$ $\sim$1$-$3 $\times$ 10$^{21}$ cm$^{-2}$.  The RS+PL model also
requires an additional absorbing column for M33 and NGC~1313
($\sim$5$-$10 $\times$ 10$^{21}$ cm$^{-2}$).  For NGC~5408, the best fit
external absorbing column is 0, although it is poorly constrained and
may be as high as 0.4 $\times$ 10$^{21}$ cm$^{-2}$ (90\% confidence
limit).

For the elliptical galaxies, the best fit model (RS+PL) also requires an
additional absorbing column over the Galactic value (for the power-law
component).  The best fit
values range from $\sim$3$-$7 $\times$ 10$^{21}$ cm$^{-2}$.

Note that for the BM fits to the spiral galaxies, essentially no intrinsic 
absorbing column is required, unlike the fits including DBB or RS components.
We attribute this to the fact that the BM model uses a single blackbody
to model the disk while the disk is usually modelled as a multiple-component
blackbody (DBB model).  Since a single blackbody model turns over at low
energies, it mimics a DBB model with absorption at low energies and less
absorption will be required.

\section{Discussion}

In order to investigate the nature of accreting BHs in nearby galaxy nuclei,
we have searched for compact X-ray sources using ROSAT HRI 
and ASCA observations
of a distance-limited 
sample of nearby galaxies.  
Before discussing the origin of the compact X-ray sources in the
elliptical and spiral galaxies, we list some possible explanations for
what the sources may be.



\subsection{Possible Explanations for the Compact X-ray Sources}

%
The compact X-ray sources could be AGN-like
sources that for some reason do not reside in the nucleus, 
very luminous 
XRBs that are by chance located near the nucleus, 
superluminal X-ray transients, bright young X-ray supernovae,
or some new,
unexplained type of X-ray source.

\subsubsection{Low-luminosity AGNs}

As noted in section 4.4, the ROSAT soft X-ray luminosities of the
compact X-ray sources are several orders of magnitude less luminous than
typical Seyfert galaxies.  Thus, if the sources are AGNs, one would
expect this difference to be due to either a lower BH mass 
(e.g., if the BH mass scales with X-ray luminosity, M$_{BH}$ may be as
low as $\sim$ 10$^4$ M$_\odot$)
or a lower
accretion rate.  If the mass accretion rate is low enough, the standard
thin disk model for accretion (Shakura \& Sunyaev 1973) is not required
and the ADAFs become possible solutions (Narayan \& Yi 1994).


Lasota et al. (1996) have used an ADAF model to explain the spectrum of
the low-luminosity AGN NGC~4258.  In their model, the outer disk is a
standard optically thick thin accretion disk, but at smaller radii their
model consists of an advection-dominated flow.  It is consistent that
mass accretion rates are low in low-luminosity AGNs.  For example, 
L/L$_{Edd}$ is estimated to be $\sim$10$^{-4}$ for the LINER galaxy
NGC~4258 (Lasota et al. 1996), when nominal values for classical AGNs
are taken to be $\sim$10$^{-3}$$-$10$^{-2}$ (e.g., Wandel 1991).

\subsubsection{Black Hole X-ray Binaries}

Another possibility is that the compact X-ray sources are extraluminous
X-ray binaries, e.g., X-ray binaries with primaries with mass
$\gapprox$10 M$_\odot$ so that the Eddington luminosities are
$\gapprox$ 10$^{39}$ erg~s$^{-1}$, which is in the range of the soft
X-ray luminosities of the compact X-ray sources in our sample of 
galaxies.
This explanation would make it easier to understand why such a large fraction
of the compact X-ray sources are displaced from the galaxy nucleus.

At some level, there is little difference between a massive BH binary
(with M$_{BH} >$ 10 M$_\odot$)
and an AGN with a low-mass (M$_{BH} \ll$ 10$^6$ M$_\odot$) BH.  Both objects
are expected to have accretion disks surrounding BHs of mass 
$\sim$ 10 $-$ 10$^4$ M$_\odot$.

ADAFs are also possible explanations for the X-ray emission in various
stages of X-ray binaries (e.g., Narayan 1997).  In low and quiescent
states, $\dot{m} < \dot{m}_{crit}$ and the inner accretion disk is
replaced with an ADAF.  If the accretion rate increases (high states), 
the standard thin disk and corona replace the ADAF and the X-ray
luminosity increases.

BH binaries are typically found in one of two states (in general terms).
In the first (soft, or high) state, the spectrum consists of a 
relatively steep ($\Gamma \approx$ 2.5) power-law at high 
(E $\gapprox$ 3 keV) energies, and a soft ``blackbody'' like component
seen below $\sim$2 keV.  BH binaries in their hard (low) state have 
power-law spectra with shallower slopes ($\Gamma \approx$ 1.3$-$1.9)
and no soft component.
As noted in section 5, the spectral slopes of the objects in the ASCA
sample are more consistent with BH X-ray binaries in their soft state.
Note that since we have a flux-limited sample, we will preferentially
select X-ray binaries which are in their most luminous state (the
high/soft state).

\subsubsection{Superluminal X-ray Sources}

We note that there are two recently-discovered BH candidates in our own
galaxy, GRS 1915+105 and GRO J1655+40, which show evidence of bipolar
superluminal outflows in radio images.
Although their X-ray luminosities are not much more than a few 10$^{38}$
erg~s$^{-1}$,
it is possible that if the sources were beamed into the line of sight,
luminosities $\gapprox$10$^{39}$ erg~s$^{-1}$ (which is comparable to
the luminosities we find for our compact X-ray sources) may be attained.
An argument against these sources being responsible for most of the
compact X-ray sources is that they are highly variable 
(e.g., Belloni et al. 1997), and it is not
clear how frequently the sources would be in their ``high'' state.

Continuous observations with BATSE in the 20~keV $-$ 2~MeV energy range
show the superluminal X-ray transient GRS~1915$+$105 to be in a ``high''
state for periods up to one year in duration (Harmon et al. 1997).
ROSAT and ASCA observations of these sources are, however, not as
complete (e.g Greiner 1996; Zhang et al. 1997), so the long-term
behavior of the soft X-ray emission is not as well known.  Based on the
BATSE levels, the flux seems to drop by a factor of up to 50 in
the ``low'' state.
If the objects happen to be superluminal sources, then one of the
expectations is that the X-ray luminosity will be quite variable and the
X-ray sources would not necessarily be located near the galaxy nucleus.

\subsubsection{Young X-ray Supernovae}

Although not much is known about how common these objects are, they
typically emit with ROSAT X-ray luminosities up to $\sim$10$^{40}$
erg~s$^{-1}$ (cf. Schlegel 1995).  Schlegel, Petre \& Colbert (1996)
monitored the X-ray bright supernova SN~1978K over a period of $\sim$4
years and found it to be constant, even though models (e.g. Chevalier
\& Fransson 1994) had predicted the light curve to drop in that amount
of time.  If these objects are indeed common and stay X-ray luminous for
long periods of time, then they might be considered possible candidates
for the compact X-ray sources we find in our sample of galaxies.  They
would presumably often be off-nuclear sources since most detected X-ray
supernovae are of type~II (cf. Schlegel 1995), which 
would tend to be found in star-forming regions located throughout the
galaxy.

\subsection{Statistical Results from the HRI Survey}

\subsubsection{Sources with Previous Evidence for Nuclear Activity 
or Central Black Holes}

Several of the elliptical galaxies in the sample are also radio galaxies
with large-scale jets, indicating the presence of an AGN.  The galaxies
of this type are NGC~4278 (Wrobel \& Heeschen 1984), NGC~4374 (M84;
Bridle \& Perley 1984), and NGC~4552 (Wrobel \& Heeschen 1984).
We find compact X-ray sources
in HRI images of all three of these galaxies.

Of the 39 galaxies in our sample, four galaxies
(NGC~221 [M32], NGC~4278, NGC~4374 [M84], and NGC~4552 [M89]) have
previous evidence for a central dark core of mass 10$^{6.5}$, 10$^{9.2}$,
10$^{9.2}$, and 10$^{8.7}$ M$_\odot$, respectively, from dynamical
modelling of the gas or stars in the nuclear region.  It is noteworthy
that in all four of these galaxies, we found compact X-ray sources,
which is highly suggestive that the X-ray emission from the sources in
these galaxies comes from accretion onto a central supermassive BH.
In three of these four galaxies, the compact X-ray source is coincident
(offset $<$ 10") with the galaxy nuclei.

Thirteen of the 39 galaxies in our sample overlap with the large sample
of nearby galaxies studied by HFS95. Six of
these galaxies have been classified by these authors as having
Seyfert-like, LINER-like  or ``transition''(between LINER and H~II
region nuclei)  nuclear optical spectra (see Table \ref{tabclass}).  Of 
these six galaxies, in all but
one (NGC~4138, classified S1.9) galaxy, we found a compact X-ray source.
In addition, in all five of these galaxies, the compact X-ray source is
coincident with the galaxy nuclei.

\placetable{tabclass}

This evidence together suggests that many of the compact X-ray sources
in our sample are caused by accretion processes onto supermassive
BHs, similar to the
processes involved in AGNs.  However, as noted by Petre et al. (1993),
in the LINER galaxy M81 there is no
evidence for an accretion disk.  
This may also be the case for the other AGN-like sources in our sample.
Since ADAF models do not involve thin
accretion disks, they make an obvious choice for the physical process
involved in explaining the X-ray emission from these AGN-like compact X-ray 
sources in our sample.

\subsubsection{Sources with No Previous Evidence for Nuclear Activity}

Of the remaining 33 galaxies {\it not} known to have Seyfert-like or
LINER-like optical nuclear spectra, we find compact near-nuclear
X-ray sources in
nearly half (15) of them.  It is these 33 galaxies that make up a sample of
truly ``normal'' galaxies and our detection frequency of $\sim$50\% 
indicates that luminous compact X-ray sources are quite common in normal
galaxies.

Note that of the 9 galaxies with offsets $>$10", all are classified as
either H~II region galaxies or are unclassified or normal galaxies
(i.e., none are classified as having Seyfert-like or LINER-like nuclei).

There are 16 galaxies with normal (or H~II region) nuclei in which we
find compact X-ray sources.  Nine (56\%) of these 16 sources
are located $>$10" from the galaxy nucleus, a far greater
percentage than for the AGN-like galaxies (0\%; 0 of 5 galaxies).

One might first suggest that many of these sources are merely X-ray
binaries that are located near the galaxy nucleus, as has been argued
for M32 (Loewenstein et al. 1998) and M33 (Dubus et al. 1997).  The
X-ray luminosities of the normal galaxy sources are significantly
lower than for the AGN-like sources (mean luminosity 2.4 $\times$
10$^{39}$ erg~s$^{-1}$ {\it vs.} 4.4 $\times$ 10$^{39}$ erg~s$^{-1}$),
also suggestive that there these are two different classes of sources.

As discussed in section 4.5, we found no noticeable difference in
luminosity between X-ray sources with large offsets and those with small
offsets.  However, as can be seen in Figure 4, those galaxies with
previous evidence for nuclear activity (filled rectangles) have smaller
separations and higher luminosities.

\subsubsection{Other Properties}

There does not seem to be any correlation between the size of the galaxy
(e.g., dwarf or normal-size galaxy) and the offset of the central X-ray
source.  We also note that the presence of H~II regions in the nuclei 
of the galaxies does not seem to be correlated with the offset of the 
X-ray sources.

\subsection{The L$_X$ vs. L$_B$ Relationship}

As discussed by Canizares, Fabbiano \& Trinchieri (1987), discrete X-ray
sources in external galaxies may make a significant contribution to
their total X-ray luminosity.  These authors find a simple relationship
between the blue luminosity from stars and the predicted X-ray emission from 
X-ray binaries.  In Figure \ref{fig5}, 
we show a plot of the ROSAT 0.2$-$2.4 keV
luminosities of nuclear pointlike sources in
the elliptical, spiral, irregular and dwarf galaxies in
our sample and compare it to the predicted X-ray luminosities from X-ray
binaries from the Canizares et al. relationship.  The blue luminosities
for the galaxies have been corrected to the 15$"$ radius aperture 
used for the X-ray luminosities, using the blue fluxes in Table \ref{tabros39}
and the correction factors listed in Table \ref{tabrosdetect}.  We used
the digital sky B-band survey plates to estimate these correction
factors.  We should note that
the relationship between L$_X$ and L$_B$ used by Canizares et al. is
based on the observed luminosities of globular clusters and of sources
in the bulge of M31.
We have included a correction from the 
0.5$-$4.5 keV band to the 0.2$-$2.4 keV band assuming a power-law
spectrum with photon index $\Gamma =$ 1.7.  

Note that all of the galaxies in our sample
have X-ray luminosities in excess (by up to three orders of magnitude) of that
predicted for X-ray binaries.  The conclusion is that, in this 15$"$
radius region, the X-ray to blue luminosity ratio is up to 10$^3$ times
larger than that found for normal galaxies (Canizares et a. 1987).  
It is highly suggestive that the X-ray
emission is not from normal X-ray binaries, but is from other sources, such as
superluminous X-ray binaries, low-luminosity AGN or X-ray bright
supernovae.

\placefigure{fig5}

\subsection{Nature of the Elliptical Galaxy X-ray Sources}


There are four elliptical galaxies in our sample (NGC~4278, NGC~4374,
NGC~4406 and NGC~4552).  We have chosen to discuss the results from
these galaxies separately since it is known that these galaxies have
extended emission from hot gas (Forman et al. 1985) as well as 
other components, such as the compact nuclear X-ray sources under study.
Matsushita et al. (1994) discovered a hard X-ray component in ASCA
spectra of five early-type galaxies in addition to the hot gas component
dominating at soft ($\lapprox$2 keV) energies.  As noted by Matsumoto et
al. (1997), the X-ray luminosity of the hard component in a sample of 12
early-type galaxies exceeds that predicted for the discrete X-ray binary
population (as estimated from the blue luminosity, Canizares et al.
1987) for three of the 12 galaxies.  Matsumoto et al. explain this
excess as probably due to X-ray emission from AGNs in these galaxies.

Referring back to Figure \ref{fig5}, we note that all of the elliptical
galaxies in our sample have ROSAT luminosities (corresponding with the
``nuclear'' source) in excess of that predicted by an X-ray binary
component.
This excess could be  a hard component from an AGN, or a soft thermal 
component localized within the HRI point-spread function.

The three elliptical galaxies 
NGC~4374, NGC~4406 and NGC~4552
have extended thermal emission within the
15\asec\ radius used to extract counts for the quoted ROSAT luminosity.
Using the correction factors for the point-like component from section
3.1, we compared the point-like component of the 0.2$-$2.4 keV ROSAT
count rate to that predicted from the hard (power-law) component of the
ASCA model (which includes flux from a much larger region).  
Note that in none of these three objects is the hard component resolved
by ASCA.  The hard (power-law) component was a factor of 3.1, 12.5 and 1.2
times larger than the ROSAT point-like component (for galaxies
NGC~4374, NGC~4406 and NGC~4552, respectively).  This indicates that the
hard component may be extended in some cases (e.g. Matsushita et al.
1994).  Another possibility is that the hard (or point-like) component
varied in intensity between the ROSAT HRI and ASCA observations.

We next turn to the results from the ASCA spectral fitting of the
elliptical galaxies NGC~4374, NGC~4406 and NGC~4552.
%
Two (NGC~4374 and NGC~4552) galaxies are also radio galaxies and
have dynamical evidence for a central dark core of mass $\sim$10$^9$
M$_\odot$.  The best fit model for these three sources was a 
Raymond-Smith plasma plus a power-law model, with the power-law 
slopes $\Gamma \sim$ 2.0$-$3.4 and thermal plasma temperatures
0.7$-$0.8 keV.  Note that these three galaxies are in dense environs.
The temperatures are consistent with hot gas known 
to be present in elliptical galaxies, but the power-law slopes,
if from AGN with M(BH) $\sim$ 10$^9$ M$_\odot$, imply a different
accretion process than typical Seyfert 1 galaxies, since their 
slopes are steeper than those seen in Seyferts by $\gapprox$ 1.

The unabsorbed X-ray luminosities of the hard power-law components are 
$\sim$2 $\times$ 10$^{40}$ erg~s$^{-1}$ in the 2$-$10 keV band, for
each of the three elliptical galaxies (see Table \ref{tabascalum}).  
This is again different (much
lower) than what is found for typical Seyfert galaxies.
Using the dynamical BH masses of $\sim$ 10$^9$ M$_\odot$ (Table 11) and
assuming L$_X$(2$-$10 keV) $\sim$ 0.1 L$_{Bol}$ (e.g., Elvis et al.
1994), this implies L/L$_{Edd}$ $\lapprox$ 10$^{-6}$, which is much
lower than estimated values of L/L$_{Edd}$ for classical AGN (cf. Wandel
1991).

\placetable{tabascalum}

Matsushita et al. (1994) 
ascribe
the hard X-ray emission in elliptical galaxy spectra 
to discrete X-ray sources in the galaxy or to
foreground/background emission from the hot intracluster medium of the
Virgo cluster. 
We find that the hard X-ray emission has a spectral slope consistent
with XRBs in their soft state and could indeed be emission from
$\sim$100 XRB (assumed to be emitting at their Eddington limit).
However, another interpretation is that the central dark core is a black
hole, and an AGN accretes gas via an accretion disk and produces 
$\sim$2 $\times$ 10$^{40}$ erg~s$^{-1}$ of X-ray emission.  
This interpretation is further supported by the high L$_X$/L$_B$ ratio
for these elliptical galaxies (section 6.3).  As mentioned
previously, narrow-line Seyfert 1 galaxies are an example of AGN with
such steep ($\Gamma \approx$ 2.5) spectra slopes.
This interpretation is supported by the fact that, as mentioned earlier,
two of the three galaxies have central dark cores with masses
$\sim$10$^9$ M$_\odot$.

Note that the RS+PL models for the elliptical galaxies yield absorbing
column  in the range 3.5$-$7.7 $\times$ 10$^{21}$ cm$^{-2}$.  These
values are slightly lower than those found in Seyfert 2 galaxies
(e.g., mean values of $\sim$ 10$^{22}$$-$10$^{23}$ cm$^{-2}$, Turner et
al. 1997), which are thought to be due to absorbing tori viewed edge-on.

\subsection{Nature of the Spiral Galaxy X-ray Sources}

Two (NGC~1313 and NGC~5408) of the three spiral galaxies 
in our ASCA sample have central X-ray
sources that are offset $\sim$1 kpc from the photometric center of the
galaxy.  In the third galaxy (M33), the central X-ray source in only
offset by $\sim$9 pc (2.6" -- not significantly different from zero).  
Such large separations might be considered
unfeasible if the X-ray sources were supermassive BHs (e.g. with 
masses $\gapprox$ 10$^6$ M$_\odot$).  However, if the objects have
masses $\lapprox$ 10$^3$ M$_\odot$, the mass of the objects 
is only a small fraction of the dynamical mass in the central kpc (which
may be $\sim$10$^7$$-$10$^8$ M$_\odot$).  So, it is not infeasible that
an object of mass $\lapprox$ 10$^3$ M$_\odot$ be found $\sim$1 kpc from
the photometric (assumed dynamical) center of the galaxy.

A list of the 2$-$10 keV unabsorbed luminosities of the X-ray sources in
the three spiral galaxies is given in Table 11 (as predicted by the
DBB+PL model).  
The masses predicted by the BM model imply Eddington ratios of
$\sim$0.04 for M33 and NGC~1313, but a slightly smaller Eddington ratio
of 0.002 for NGC~5408.  
These Eddington ratios are significantly larger than that estimated for
the  ADAF LINER galaxy NGC~4258 ($\sim$10$^{-4}$, Lasota et al. 1996).

The X-ray sources
in the spiral galaxies have 2$-$10 keV X-ray luminosities
$\sim$0.6$-$3 $\times$ 10$^{39}$ erg~s$^{-1}$, which are an order of
magnitude smaller than the hard components in the elliptical galaxies,
indicating that they may be of different origin.  These luminosities are
still one or two orders of magnitude brighter than typical BH binaries
(e.g. Tanaka \& Lewin 1995), which are though to have BHs of masses on
the order of a few solar masses.  If we roughly scale the masses to the
X-ray luminosities (M$_{BH} \sim$ 10$^6$$-$10$^9$ M$_\odot$ for L$_X
\sim$ 10$^{42}$$-$10$^{44}$ erg~s$^{-1}$), 
we arrive at masses consistent with those
predicted by the BM model ($\sim$10$^2$ $-$ 10$^6$ M$_\odot$).

Values of  r$_{in}(\cos\theta)^{{1}\over{2}}$ and kT$_{in}$ for LMXBs
are typically 5$-$10 km and 1.0$-$1.5 keV, respectively, whereas for
BHCs, the values are 20$-$30 km and 0.6$-$1.0 keV, respectively
(e.g., see Yaqoob, Ebisawa, \& Mitsuda 1993).  As mentioned in section
5, for M33 and NGC~1313\#2,
the values of r$_{in}(\cos\theta)^{{1}\over{2}}$ are much higher than
typical values for LMXBs and are more consistent with the range for
BHCs.  However, kT$_{in}$ is more consistent with the range for the
LMXBs.  We disregard the fitting results from NGC~1313\#1, since its
shallow slope suggests it is not in a soft (high) state.  
The results
for NGC~5408 are indeed much different than might be expected for a LMXB
or ``normal'' BHC.  The value of r$_{in}(\cos\theta)^{{1}\over{2}}$ is
very high (2200 $-$ 62500 km) and kT$_{in}$ is much lower than expected
for a ``normal'' BHC.  

We should note that in the standard thin accretion disk theory, r$_{in}$
extends inward to the radius of the innermost stable orbit, which is
equal to 3r$_g$ for a Schwarzschild BH.  In Table \ref{tabrincostheta},
we list the
corresponding BH mass for r$_{in}(\cos\theta)^{{1}\over{2}}$ $=$ 3r$_g$.
Note that these masses are much lower than those predicted by the 
BM model.  The discrepancy can be explained by the fact that the
multicolor disk blackbody model used by XSPEC approximates T(r)
$\propto$ r$^{-3/4}$ (1 - (6r$_g$/r)$^{{1}\over{2}}$) as T(r) $\propto$
r$^{-3/4}$, so it underpredicts the normalization and thus also the
value of r$_{in}(\cos\theta)^{{1}\over{2}}$ (L. Titarchuk, priv. comm.).
Therefore, values of r$_{in}(\cos\theta)^{{1}\over{2}}$ are only useful
for relative comparisons (e.g., to values predicted for other objects),
but not for estimates of the innermost stable orbit and thus the mass of
the BH.

It is noteworthy, however, that the predicted mass of the X-ray emitting
BH in NGC~5408 (from the BM model) is $\sim$10$^2$$-$10$^3$ times as large 
as that for M33 or NGC~1313, so it is consistent that NGC~5408 is a BHC
with a larger BH mass.

So, the
working model is that these X-ray objects are BH candidates of mass
$\sim$10$^2$ $-$ 10$^4$ M$_\odot$, currently in their soft (high) state
(except for the observation NGC~1313\#1, which is consistent with a BHC
in its hard state).

It is not clear how BHs of mass $\sim$10$^2$$-$10$^4$ M$_\odot$ would
form.  Brown \& Bethe (1994) indicated that there are two types of BHs
that are formed at the end of stellar evolution: $\sim$1.5 M$_\odot$ BHs
from supernova explosions of massive main sequence stars, and $\sim$10
M$_\odot$ BHs from the collapse of the helium core of stars more massive
than 25$-$30 M$_\odot$.  The masses implied for M33, NGC~1313 and
NGC~5408 are much higher than can be explained by either of these
processes.  

At the other extreme are supermassive BHs which are present in the nuclei
of AGNs (and possibly in the nuclei of other ``inactive'' galaxies as well).  
These BHs are assumed to be formed by the collapse of primordial gas
clouds (e.g., Silk \& Rees 1998), but formation of BHs with masses
$\lapprox$ 10$^6$ M$_\odot$ are noted to be less efficient and the
production of such BHs is likely to be inhibited.  One reason why such
objects are prevented from forming is that primordial clouds of mass
$\lapprox$10$^9$ M$_\odot$ are readily disrupted by supernova-driven
winds (Dekel \& Silk 1986).  Therefore, the BH masses of the X-ray
sources described in this paper are not likely to have been formed by
collapse of a primordial cloud.

Lee (1995) offers a possible explanation for intermediate-mass BHs, in which 
stellar-size BHs (M $\lapprox$ 10 M$_\odot$) merge and can form a seed BH
of mass $\sim$100 M$_\odot$.

We find that the DBB+PL model predicts intrinsic absorbing columns of
$\sim$1$-$3 $\times$ 10$^{21}$ cm$^{-2}$.  These columns are not typical
of the large columns through absorbing tori in Seyfert 2 galaxies.  This
implies, that if it the objects do have surrounding optically-thick tori
like those found in AGNs, then either the line-of-sight through the tori
is not edge-on, or the tori are not as optically thick as those in
AGNs.

\subsection{The Nature of the Compact X-ray Sources}

There does not seem to be a single explanation that works for all of the
compact X-ray sources.  For example, the X-ray luminosities of the
sources in the galaxies with previous evidence for nuclear activity are
significantly larger than those luminosities of the sources in the
otherwise normal galaxies.  The mean separation of the X-ray sources
from the galaxy nucleus is also larger for the normal galaxies.  This
suggests that there may be two different types of X-ray emitting objects
in these two types of galaxies.

The soft components in the ASCA spectra in the elliptical galaxies are
better fit by a Raymond-Smith thermal plasma than a multicolor disk
blackbody model.  This suggests that some of the X-ray emission from the
central X-ray sources in the elliptical galaxies may be emission from
hot gas, a component which has been known to be present in many other
elliptical galaxies (Forman et al. 1985).

What does seem to be universal among the X-ray sources is that they have
relatively steep spectral slopes compare with those of Seyfert 1 nuclei.
This, and the fact that large absorbing columns are not found, suggests
that the X-ray sources are probably different physically or
geometrically from typical AGN in active X-ray bright galaxies.

We should also mention that the preference for low absorbing columns
could also be due to a selection effect.  X-ray sources with similar
intrinsic (unabsorbed) luminosities but larger absorbing columns (e.g.,
N$_H$ $\sim$ 10$^{23}$$-$10$^{24}$ cm$^{-2}$) would be considerably
weaker ROSAT X-ray sources.

\section{Summary and Conclusions}

Using archival ROSAT HRI observations, we have used a distance-limited sample
of 39 nearby galaxies to search for compact ``nuclear'' X-ray sources.
We found compact X-ray sources within $\sim$2' of the nucleus
(photometric center) of 21 of the 39 galaxies.  Approximately 40\% (9 of
21) of the compact X-ray sources were located $>$10" ($\gapprox$ 0.3 kpc
at the mean distance of our sample) from the photometric center of the
galaxy.  

Compact X-ray sources are found in nearly all cases in which previous
possible evidence for a BH or an AGN had been found (e.g., large-scale
radio sources, dynamical dark core masses $>$ 10$^6$ M$_\odot$, or
optical spectral classification as Seyfert, LINER, or ``transition''
object).  This indicates that some of the X-ray sources in these
galaxies are probably AGN or at least accretion-driven objects with
supermassive BHs.

We analyzed archival ASCA spectra of six of the 21 galaxies with
compact X-ray sources (3 spiral galaxies and 3 elliptical galaxies).
The elliptical galaxy spectra are adequately fit with a two-component
Raymond-Smith thermal plasma model plus an absorbed power-law model.
The intrinsic absorption of the power-law source was typically several
$\times$ 10$^{21}$ cm$^{-2}$ (significantly smaller than seen in Seyfert
2 galaxies) and the power-law slope was significantly steeper ($\Gamma
\approx$ 2.5) than seen in type 1 or 2 AGNs.  The spiral galaxy spectra
are well-fit by a two-component model consisting of a multicolor disk
blackbody model plus a power-law model.  Assuming both components 
suffer the same intrinsic absorption, we find intrinsic absorption of
a few 10$^{21}$ cm$^{-2}$ and, again, steep power-law slopes of 
$\Gamma \approx$ 2.5.  The central X-ray source in the spiral galaxy 
NGC~1313 seems to have changed from a hard state to a soft state in the
2.4 intervening years between observations.  The spiral galaxy X-ray
sources, when fit with the Bulk Motion Comptonization model of Shrader
\& Titarchuk (1998), predict black hole masses $\sim$10$^2$$-$10$^4$
M$_\odot$.
This model assumes the presence of a thin disk surrounding the black
hole.

The emerging picture of the compact X-ray sources in nearby galaxies is
that they are probably high-mass (M $\gapprox$ 10 M$_\odot$) accreting
black holes.  It is not clear whether they are binary systems or not,
but the steep hard X-ray emission is consistent with that of a BHC in its
soft (high) state.  Such steep slopes are also consistent with what has
been observed in Narrow-Line Seyfert 1 galaxies.  

The X-ray sources in the spiral galaxies seem to have BHs of
intermediate mass ($\sim$10$^2$$-$10$^4$ M$_\odot$), which make them
members of a unique class of accreting black holes.  
BHs of this mass range are not predicted in most theories of BH
formation.
Except for the central X-ray source in M33, 
there are not any such
objects in the Local Group (M31, the Milky Way, SMC, LMC)
that are known to emit at intermediate X-ray
luminosities of L$_X \sim$ 10$^{39}$$-$10$^{40}$ erg~s$^{-1}$.
Future studies of
these interesting objects in nearby galaxies are imperative for
determining their exact nature.

\acknowledgments

EJMC would like to thank Lev Titarchuk for many helpful discussions and
for calculations having to do with the bulk motion model.  EJMC
acknowledges support from the National Research Council.  This research
has made extensive use of the NASA/IPAC Extragalactic Database (NED),
which is operated by the Jet Propulsion Laboratory, Caltech, under
contract with NASA.  The Digital Sky Survey data were retrieved from the
STScI WWW page and were produced under US government grant NAGW-2166.

\clearpage

 
\begin{deluxetable}{llrrrrcrrrr}
\scriptsize
\tablecaption{Sample of Nearby Galaxies \label{tabros39} }
\tablewidth{0pt}
\tablehead{
\colhead{Galaxy} & \colhead{Other} & \colhead{R.A.\tablenotemark{a}} & 
\colhead{Dec.\tablenotemark{a}} & 
\colhead{$cz$\tablenotemark{a}} & \colhead{D\tablenotemark{b}} & 
\colhead{ref.\tablenotemark{b}} &
\colhead{Gal.} & 
\colhead{Log } & \colhead{f$_B$\tablenotemark{c}} & 
\colhead{f$_{FIR}$\tablenotemark{d}} \\
\colhead{Name} & \colhead{Name} & \multicolumn{2}{c}{(J2000)} & 
\colhead{($km$ } & \colhead{} & \colhead{} &
\colhead{Type\tablenotemark{a}} &
\colhead{Axial} & \colhead{(log $erg$} &  \colhead{(log $erg$ } \\
\colhead{} & \colhead{} & \colhead{} & \colhead{} &
\colhead{$s^{-1}$)} & 
\colhead{(Mpc)} &
\colhead{} & \colhead{} & \colhead{Ratio\tablenotemark{a}} &
\colhead{ $s^{-1}~cm^{-2}$)} & \colhead{ $s^{-1}~cm^{-2}$)} \\
} 
\startdata
NGC~205  & M110 & 0 40 22.5 & $+$41 41 11 &  $-$232 & 0.7 & 1 & $-$5.0 & 0.30 &
  $-$8.70 & -10.17 \\
NGC~221  & M32  & 0 42 41.9 & $+$40 51 55 &  $-$205 & 0.7 & 1 & $-$6.0 & 0.13 &
 $-$8.68 & {...} \\
NGC~300  &       & 0 54 53.8 & $-$37 40 57 &  142 &   1.2 & 1 &  7.0 & 0.15 &
 $-$8.58 & $-$8.77 \\
NGC~598  &  M33  & 1 33 50.9 & $+$30 39 37 &  $-$179 &  0.7 & 1 & 6.0 &  0.23 &
 $-$7.49 & $<-$9.04 \\
NGC~1313 &       & 3 18 15.5 & $-$66 29 51 &  457 &  3.7 & 1 & 7.0 & 0.12 &
 $-$8.90 & $-$9.03 \\
NGC~1396  &      & 3 38 06.3 & $-$35 26 25 &   894 & 16.9 & 2 & $-$3.0 & 0.07 &
 $-$11.10 & {...} \\
AM~0337-353 &    &  3 39 13.6 & $-$35 22 19 &   952 & 16.9 & 2 & $-$2.0 & 0.10 &
 $-$11.30 & {...} \\
IC~342     &     & 3 46 49.7 & $+$68  5 45  &   34 &  3.9 & 1 & 6.0 & 0.01 &
 $-$7.42 & $-$8.36 \\
NGC~1705  &   &    4 54 13.6 & $-$53 21 44  &  673 & 6.0 & 1 & $-$3.0 & 0.13 &
 $-$10.22 & $-$10.29 \\
NGC~2403  &    &   7 36 54.5 & $+$65 35 58  &  131 & 4.2 & 1 &  6.0 & 0.25 &
 $-$8.56 & $-$8.85 \\
Ho~II &        &   8 19 06.0 &  $+$70 42 51 &  158 & 1.4 & 3 & 10.0 & 0.10 &
 $-$9.54 & $-$10.15 \\
ESO~495-G21 & He~2-10 & 8 36 15.0 &  $-$26 24 34 &  896 & 6 & 4 & 90.0 & 0.11 &
 {...} & {...} \\
UGC~5086 &     & 9 32 54 & $+$21 29  0  & 448 & 2 & 5 & 10.0 & 0.00 &
 {...} & {...} \\
NGC~3031  & M81 & 9 55 33.5 & $+$69  4 00  &  -34 &  1.4 & 1 & 2.0 & 0.28 &
 $-$8.14 & $-$9.20 \\
Ho~IX &   &       9 57 30  & +69  2  0  & 46 & 1.4 & 3 & 10.0 & 0.09 &
 $-$10.82 & {...} \\
Leo~A &   &     9 59 18  &  +30 44  0  &  20 &  6 & 5 & 10.0  & 0.22 &
 $-$10.26 & {...} \\
NGC~3990 &  &   11 57 36.3 &  $+$55 27 33 &   696 & 17.0 & 1 & -3.0 & 0.24 &
 $-$10.57 & {...} \\
NGC~4138  &  &   12  9 30.8 &  $+$43 41 16 &   923 & 17.0 & 1 & $-$1.0 & 0.18 &
 $-$10.03 & {...} \\
NGC~4150     & & 12 10 33.5 & $+$30 24 06  &  244 & 9.7 & 1 & $-$2.0 & 0.16 &
 $-$10.15 & $-$10.16 \\
NGC~4190    & &  12 13 44.7 &  $+$36 38 00  &  230 & 2.8 & 1 & 10.0 & 0.04 &
 $-$10.53 & $-$10.57 \\
UGCA~276 & DDO~113 & 12 14 54 & $+$36 13 0  &  284 & 4 & 5 &  10.0 & 0.00 &
 {...} & {...} \\
UGC~7356 &      & 12 19  0 & $+$47  5  0 & 272 &  4 & 5 & 10.0 & 0.03 &
 {...} & {...} \\
NGC~4278 &     &  12 20 07.2 & $+$29 16 47 & 649 & 9.7 & 1 & $-$5.0 & 0.03 &
 $-$9.58 & $-$10.42 \\
NGC~4286 &    &   12 20 41.8 & $+$29 20 38 & 623 & 10 & 6 & 0.0  & 0.20 &
 $-$10.74 & {...} \\
\tableline
\tablebreak
NGC~4374 & M84 &     12 25 03.7 & $+$12 53 15 &   910 & 16.8 & 1 & $-$5.0 & 0.06 &
 $-$9.19 & $-$10.54 \\
IC~3303  & &     12 25 15.0 & $+$12 42 53 &  $-$213 &  16.8 & 7 &  $-$4.4 & 0.20 &
 $-$11.06 & {...} \\
NGC~4387  & &   12 25 41.8 &  $+$12 48 42 & 561 & 16.8 & 1 & $-$5.0 & 0.21 &
 $-$10.34 & {...} \\
NGC~4406  & M86 & 12 26 11.8 &  $+$12 56 49 &  $-$248 & 16.8 & 1 &  $-$5.0 & 0.19 &
 $-$9.08 & {...} \\
NGC~4449   & &  12 28 11.4 & $+$44  5 40  &   201 & 3.0 & 1 & 10.0 & 0.15 &
 $-$9.16 & {...} \\
NGC~4485  & &   12 30 31.7 & $+$41 42 01 &  493 & 9.3 & 1 & 10.0 & 0.15 &
 $-$10.08 & {...} \\
IC~3540   & &   12 35 27.3 & $+$12 45 00 &   733 & 16.8 & 7 & 11.0 & 0.10 &
 $-$11.12 & {...} \\
NGC~4552 &  M89 &  12 35 39.9 & $+$12 33 25  &  311 & 16.8 & 1 & $-$5.0 & 0.04 &
 $-$9.42 & {...} \\
NGC~4569 &  M90 &  12 36 50.1 & $+$13  9 48 & $-$235 & 16.8 & 1 & 2.0 & 0.34 &
 $-$9.10 & $-$9.26 \\
NGC~4627 &  &   12 41 59.7 & $+$32 34 29 &  765 & 7.5 & 8 & $-$5.0  & 0.16 &
 $-$10.38 & {...} \\
NGC~5204 & &    13 29 36.4 &  $+$58 25 04 &   204 & 4.8 & 1 &  9.0 & 0.22 &
 $-$9.78 & $-$9.84 \\
NGC~5408 &  &   14  4 22 &   $-$41 22 18 & 509 &  4.9 & 1 & 9.7 &  0.31 &
 $-$9.86 & $-$9.89 \\
NGC~6822 &  &  19 44 57.9 & $-$14 48 11 &  $-$56 & 0.7 & 1 &  10.0 & 0.06 &
 $-$8.54 & $-$9.40 \\
NGC~6946 &  &    20 34 52.0 &  $+$60  9 15 & 52 &  5.5 & 1 & 6.0 & 0.07 &
 $-$8.30 & $-$8.42 \\
NGC~7320 &  &   22 36 04.2 &  $+$33 56 52 &   776 &  13.8 & 1 & 7.0 & 0.29 &
 $-$10.22 & {...} \\
\enddata

 
\tablenotetext{a}{Right ascension, declination, recessional velocity,
galaxy type, and axial ratio were taken from RC3.}
\tablenotetext{b}{Distance to galaxy and reference.  References: (1)
  Tully 1988; (2) in Fornax cluster, assumed to be at distance of Fornax
  cluster (16.9 Mpc, Tully 1988); (3) in M81 group, assumed to be at
  distance of M81 (1.4 Mpc, Tully 1988); (4) Johansson 1987; (5) Melisse
  \& Israel 1994; (6) Sandage \& Hoffman 1991; (7) in Virgo cluster,
  assumed to be at distance of Virgo cluster (16.8 Mpc, Tully 1988); 
  (8) companion to NGC~4631, assumed to be at the distance to NGC~4631
  (7.5 Mpc, Hummel et al. 1984)  
   }
\tablenotetext{c}{B-band optical flux, calculated from B$^T_0$ (from RC3)}
\tablenotetext{d}{Far-infrared flux calculated using the method of Fullmer \& 
  Lonsdale 1989, using IRAS fluxes from Moshir et al. 1990 (as listed in the 
  NASA Extragalactic Database [NED])}
\end{deluxetable}

\clearpage
 
\begin{deluxetable}{lcrlcr}
\scriptsize
\tablecaption{ROSAT Observation Log \label{tabrosobs} }
\tablewidth{0pt}
\tablehead{
\colhead{Galaxy} & \colhead{Seq. No.\tablenotemark{a}} & 
\colhead{Exp. Time\tablenotemark{b}} &
\colhead{Galaxy} & \colhead{Seq. No.\tablenotemark{a}} & 
\colhead{Exp. Time\tablenotemark{b}} \\
\colhead{Name} & \colhead{} & \colhead{(ks)} &
\colhead{Name} & \colhead{} & \colhead{(ks)} \\
} 
\startdata
NGC~205  & rh600816n00 & 28.1 & UGCA~276 & rh600741n00 & 42.6 \\
NGC~221  & rh600600n00 & 12.7 & UGC~7356 & rh701008n00 & 27.6 \\
NGC~300  & rh600621a01 & 19.1 & NGC~4278 & rh701004a01 &  5.9 \\
NGC~598  & rh400460a01 & 41.0 & NGC~4286 & rh701004a01 &  5.9 \\
NGC~1313 & rh500403a01 & 31.4 & NGC~4374 & rh600493n00 & 26.5 \\
NGC~1396 & rh600831n00 & 73.5 & IC~3303  & rh600493n00 & 26.5 \\
AM~0337-353 & rh600831n00 & 73.5 & NGC~4387 & rh600493n00 & 26.5 \\
IC~342   & rh600022n00 & 19.1 & NGC~4406 & rh600463a02 & 10.5 \\
NGC~1705 & rh600604n00 & 26.2 & NGC~4449 & rh600743n00 & 15.4 \\
NGC~2403 & rh600767n00 & 26.5 & NGC~4485 & rh600697n00 &  6.7 \\
Ho~II    & rh600745n00 &  7.8 & IC~3540  & rh600491a01 & 30.1 \\
ESO~495-G21 & rh600700n00 & 20.3 & NGC~4552 & rh600491a01 & 30.1 \\
UGC~5086 & rh600602n00 & 13.6 & NGC~4569 & rh600603a01 & 21.9 \\
NGC~3031 & rh600247n00 & 26.6 & NGC~4627 & rh600193a01 & 14.3 \\
Ho~IX    & rh600247n00 & 26.6 & NGC~5204 & rh702723n00 & 14.9 \\
Leo~A    & rh600117n00 & 19.1 & NGC~5408 & rh600376n00 & 12.6 \\
NGC~3990 & rh600430n00 & 13.6 & NGC~6822 & rh600815a01 & 52.9 \\
NGC~4138 & rh701943n00 &  5.8 & NGC~6946 & rh600501n00 & 60.3 \\
NGC~4150 & rh600762a01 & 10.5 & NGC~7320 & rh800689n00 & 23.3 \\
NGC~4190 & rh600733n00 &  3.5 &          &             & \\
\enddata

 
\tablenotetext{a}{ROSAT HRI sequence number for archival data}
\tablenotetext{b}{Exposure time in kiloseconds (Keyword EXPTIME in the FITS 
      headers for the data)}
\end{deluxetable}


\clearpage
\begin{deluxetable}{lllclrr}
\scriptsize
\tablecaption{ASCA Observations of Nearby Galaxies \label{tabascaobs}}
\tablewidth{0pt}
\tablehead{
\colhead{Galaxy} & 
\colhead{Seq. No.} & 
\colhead{Obs. Date} &
\colhead{CCD mode} &
\colhead{SIS0 cnt. rate\tablenotemark{a}} &
\colhead{SIS0 on-time} &
\colhead{GIS2 on-time} \\
\colhead{Name} & 
\colhead{} &
\colhead{(YYYYMMDD)} &
\colhead{} &
\colhead{(cnt~s$^{-1}$)} &
\colhead{ksec} &
\colhead{ksec} \\
} 
\startdata
M33 (\#1) & 60036000 & 19930722$-$19930723 & 4 & 0.45 & 13.8 & 17.6 \\
M33 (\#2) & 60036010 & 19930723            & 4 & 0.46 & 12.9 & 18.3 \\
NGC~1313 (\#1) & 60028000 & 19930712$-$19930713 & 4 & 0.066 & 18.2 & 29.8 \\
NGC~1313 (\#2) & 93010000 & 19951129$-$19951130 & 1 & 0.051 & 30.4 & 36.3 \\
NGC~4374 & 60007000 & 19930704 & 2 & 0.071 & 14.0 & 21.1 \\
NGC~4406 & 60006000 & 19930703$-$19930704 & 2 & 0.29 & 5.7 & 20.9 \\
NGC~4552 & 63023000 & 19950626$-$19950627 & 2 & 0.032 & 27.6 & 27.8 \\
NGC~5408 & 93011000 & 19950208$-$19950209 & 2 & 0.048 & 44.6 & 48.5 \\
\enddata
 
\tablenotetext{a}{Count rate in the SIS0 detector for the energy range
0.4$-$10 keV}
\end{deluxetable}


\clearpage
\begin{deluxetable}{lrrrc}
\scriptsize
\tablecaption{Galaxies without Detected Nuclear X-ray Sources
  \label{tabrosundetect} }
\tablewidth{0pt}
\tablehead{
\colhead{Galaxy} & 
\colhead{Cnt Rate\tablenotemark{a}} & \colhead{N$_H^{Gal}$$^{,b}$} &
\colhead{f$_{X}$} &
\colhead{L$_{X}$} \\
\colhead{Name} & 
\colhead{10$^{-2}$ s$^{-1}$} & 
\colhead{10$^{20}$ cm$^{-2}$} &
\colhead{10$^{-13}$ cgs} &
\colhead{10$^{38}$ cgs} \\
} 
\startdata
NGC~205     & $<$ 0.47 &  6.75  &  $<$ 2.633  & $<$ 0.154 \\
NGC~300     & $<$ 0.24 &  3.23  &  $<$ 1.119  & $<$ 0.193 \\
NGC~1396    & $<$ 0.41 &  1.35  &  $<$ 1.573  & $<$ 53.8 \\
AM~0337-353 & $<$ 0.31 &  1.30  &  $<$ 1.180  & $<$ 40.3 \\
NGC~1705    & $<$ 0.36 &  4.19  &  $<$ 1.787  & $<$ 7.70 \\
UGC~5086    & $<$ 0.24 &  3.08  &  $<$ 1.106  & $<$ 0.529 \\
Ho~IX       & $<$ 0.31 &  4.07  &  $<$ 1.528  & $<$ 0.358 \\
Leo~A       & $<$ 0.32 &  1.68  &  $<$ 1.285  & $<$ 5.54 \\
NGC~3990    & $<$ 0.24 &  1.21  &  $<$ 0.9010 & $<$ 31.2 \\
NGC~4138    & $<$ 0.68 &  1.36  &  $<$ 2.613  & $<$ 90.4 \\
UGCA~276    & $<$ 0.33 &  1.49  &  $<$ 1.292  & $<$ 2.47 \\
UGC~7356    & $<$ 0.31 &  1.16  &  $<$ 1.154  & $<$ 2.21 \\
NGC~4286    & $<$ 0.15 &  1.79  &  $<$ 0.6106 & $<$ 7.31 \\
IC~3303     & $<$ 0.43 &  2.58  &  $<$ 1.901  & $<$ 64.2 \\
NGC~4387    & $<$ 0.30 &  2.60  &  $<$ 1.329  & $<$ 44.9 \\
IC~3540     & $<$ 0.27 &  2.55  &  $<$ 1.191  & $<$ 40.2 \\
NGC~4627    & $<$ 0.27 &  1.27  &  $<$ 1.023  & $<$ 6.89 \\
NGC~7320    & $<$ 0.26 &  7.96  &  $<$ 0.9878 & $<$ 22.5 \\
\enddata
 
\tablenotetext{a}{3$\sigma$ upper limit to the count rate for a 15"
radius circle detect cell}
\tablenotetext{b}{Galactic absorbing column calculated with HEASARC's
on-line Galactic hydrogen database (data from Dickey \& Lockman 1990)}
\end{deluxetable}


\clearpage
 
\begin{deluxetable}{lrrrrrrrrr}
\scriptsize
\tablecaption{Compact X-ray Sources in Nuclear Regions of Nearby Galaxies
   \label{tabrosdetect} }
\tablewidth{0pt}
\tablehead{
\colhead{Galaxy} & \multicolumn{2}{c}{X-ray Position (J2000)} & 
\multicolumn{2}{c}{Offset\tablenotemark{b}} &
\colhead{Cnt Rate\tablenotemark{c}} & \colhead{N$_H^{Gal}$$^{,d}$} & 
\colhead{f$_{X}$\tablenotemark{e}} &
\colhead{L$_{X}$\tablenotemark{e}} &
\colhead{$log(f_B)$\tablenotemark{f}} \\
\colhead{Name} & 
\colhead{R.A.\tablenotemark{a}} & \colhead{Dec.\tablenotemark{a}} & 
\colhead{(arcsec)} & \colhead{(pc)} & 
\colhead{10$^{-2}$ s$^{-1}$} & 
\colhead{10$^{20}$ cm$^{-2}$} &
\colhead{10$^{-13}$ $cgs$} &   
\colhead{10$^{38}$ $erg~s^{-1}$} &
\colhead{} \\
} 
\startdata
M32 &         00 42 42.7 & $+$40 51 51.1 & 9.9 & 33.6 & 1.19 & 6.34 & 6.556 &  0.384 & -0.45\\
M33 &         01 33 51.1 & $+$30 39 37.0 & 2.6 & 8.8 & 7.23\tablenotemark{g} & 5.58 & 38.53\tablenotemark{g} & 2.26\tablenotemark{g} & -2.98 \\
NGC~1313  &   03 18 20.3 & $-$66 29 11.3 & 49.0 & 879 & 3.01 & 3.90 & 14.68 & 24.1 & -1.94 \\
IC~342 &      03 46 48.6 & $+$68 05 51.0 & 8.6 &  163 & 0.65 & 30.2 & 6.603 & 12.0 & -2.08 \\
NGC~2403  &   07 36 55.7 & $+$65 35 42.6 & 17.1 & 348 & 0.36 & 4.11 & 1.779 & 3.76 & -2.07 \\
Ho~II     &   08 19 29.7 & $+$70 42 18.4  & 121.9 & 827 & 11.99 & 3.42 & 56.65 & 13.3 & -1.74 \\
ESO~495-G21 & 08 36 15.3 & $-$26 24 32.7 & 4.2 & 122 & 0.47 & 9.84 & 2.928 & 38.6 & {...} \\
M81        &  09 55 33.4 & $+$69 03 53.1 & 6.9 & 46.8 & 11.99\tablenotemark{g} & 4.16 & 5.940\tablenotemark{g} & 1.39\tablenotemark{g} & -2.54 \\
NGC~4150   &  12 10 34.7 & $+$30 24 00.9 & 16.3 & 767 & 2.23 &  1.62 & 8.886 & 100.0 &  -0.69 \\
NGC~4190   &  12 13 45.3 & $+$36 37 58.5 & 7.4 & 100 & 2.35 & 1.54 & 9.266 &  8.69 & -0.71 \\
NGC~4278   &  12 20 06.7 & $+$29 16 50.4 & 7.4 & 348 & 1.72 & 1.77 & 6.987 &  78.7 & -1.12 \\
NGC~4374   &  12 25 03.7 & $+$12 53 13.5 & 1.5 & 122 & 0.29\tablenotemark{g} & 2.60 & 1.278\tablenotemark{g} & 43.2\tablenotemark{g} & -1.55 \\
NGC~4406   &  12 26 11.7 & $+$12 56 46.7 & 2.7 & 220 & 0.30\tablenotemark{g} & 2.62 & 1.339\tablenotemark{g} & 45.3\tablenotemark{g} & -1.74 \\
NGC~4449   &  12 28 09.4 & $+$44 05 07.1 & 39.3 & 572 & 0.57 & 1.37 & 1.940 & 2.09 & -1.41 \\
NGC~4485   &  12 30 30.5 & $+$41 41 45.5 & 20.5 & 924 & 0.39 & 1.78 & 1.586 & 16.4 & -0.79 \\
NGC~4552   &  12 35 39.7 & $+$12 33 28.6 &  4.6 & 375 & 0.28\tablenotemark{g} & 2.57 & 1.232\tablenotemark{g} & 41.6\tablenotemark{g} & -1.21 \\
NGC~4569   &  12 36 49.7 & $+$13 09 47.2 &  5.9 & 481 & 0.46 & 2.49 & 1.654 & 55.9 & -1.45 \\
NGC~5204   &  13 29 38.6 & $+$58 25 03.2 & 17.3 & 403 & 2.50 & 1.39 & 9.649 & 26.6 & -1.09 \\
NGC~5408   &  14 03 19.4 & $-$41 22 57.7 & 49.3 & 1171 & 4.88 & 5.66 & 26.11 & 75.0 & -0.90 \\
NGC~6822   &  19 44 56.4 & $-$14 48 29.3 & 28.4 & 96.4 & 0.35 & 9.49 & 1.334 & 0.0782 & -2.21 \\
NGC~6946   &  20 34 52.4 & $+$60 09 09.9 &  5.9 & 157 & 0.24 & 21.1 & 2.003 & 7.25 & -1.91 \\
\enddata

 
\tablenotetext{a}{X-ray position of compact source.  The uncertainties in
  the X-ray positions are $\sim$10$"$ due to pointing uncertainties of the
  ROSAT spacecraft.}
\tablenotetext{b}{Offset of X-ray position from optical photometric
  center (Table 1)}
\tablenotetext{c}{HRI count rate using source region of 15" radius
  circle}
\tablenotetext{d}{Galactic absorption column (Dickey \& Lockman 1990)}
\tablenotetext{e}{0.2$-$2.4 keV X-ray flux (units $erg~s^{-1}~cm^{-2}$)
  and luminosity, assuming a
  power-law spectrum (see text) and absorption column listed in the Table.
  Distances listed in Table 1 were used to compute luminosities.}
\tablenotetext{f}{Logarithm of the fraction of blue light in the 15$"$
  radius aperture centered on the X-ray position (columns 2 and 3)
  compared with the blue light from the whole galaxy.  See section 6.3.}
\tablenotetext{g}{HRI count rates, fluxes and luminosities have been
  corrected to reflect the contribution from the point-like component (see
  section 3.1)}
\end{deluxetable}


\clearpage
\begin{deluxetable}{lrccccr}
\scriptsize
\tablecaption{Single-Component Fits to ASCA Data \label{tabascasingle} }
\tablewidth{0pt}
\tablehead{
\colhead{Galaxy} & 
\colhead{Model\tablenotemark{a}} &
\colhead{$\Gamma$ or kT (keV)\tablenotemark{b}} &
\colhead{Normalization\tablenotemark{c}} &
\colhead{N$_H$\tablenotemark{d}} &
\colhead{Abundance\tablenotemark{e}} &
\colhead{$\chi^2$ / d.o.f.\tablenotemark{f}} \\
\colhead{Name} & 
\colhead{} &
\colhead{} &
\colhead{} &
\colhead{(10$^{21}$ cm$^{-2}$)} &
\colhead{} &
\colhead{} \\
} 
\startdata
\multicolumn{3}{l}{Spiral Galaxies}
\\
M33 (\#1) & PL & $\Gamma =$ 2.33 & 6.95 $\times$ 10$^{-3}$ & 3.20 & & 659 / 595 \\
         &    &  2.28 $-$ 2.37  &                         & 2.94 $-$ 3.47 & & \\
         & RS & kT $=$ 3.33     & 1.88 $\times$ 10$^{-2}$ & 1.45 & 0 & 555 / 594 \\
         &    &  3.17 $-$ 3.48  &                         & 1.28 $-$ 1.64 & $<$ 0.05 & \\
M33 (\#2) & PL & $\Gamma =$ 2.31 & 7.42 $\times$ 10$^{-3}$ & 2.99 & & 838 / 676 \\
         &    &  2.27 $-$ 2.35  &                         & 2.77 $-$ 3.21 & & \\
         & RS & kT $=$ 3.43     & 2.00 $\times$ 10$^{-2}$ & 1.27 & 0 & 652 / 675 \\
         &    &  3.31 $-$ 3.56  &                         & 1.14 $-$ 1.41 & $<$ 0.04 & \\
NGC~1313 (\#1) & PL & $\Gamma =$ 1.74 & 4.74 $\times$ 10$^{-4}$ & 0.151 & & 244 / 259 \\
              &    &  1.67 $-$ 1.82  &                         & $<$ 0.57 & & \\
              & RS & kT $=$ 6.68     & 1.77 $\times$ 10$^{-3}$ & 0 & 0.10 & 253 / 258 \\
              &    &  5.92 $-$ 7.47  &                         & $<$ 0.09 & $<$ 0.27 & \\
NGC~1313 (\#2) & PL & $\Gamma =$ 2.77 & 1.27 $\times$ 10$^{-3}$ & 2.95 & & 185 / 198 \\
              &    &  2.65 $-$ 2.88  &                         & 2.50 $-$ 3.43 & & \\
              & RS & kT $=$ 2.05     & 3.16 $\times$ 10$^{-3}$ & 0.910 & 0.06 & 196 / 197 \\
              &    &  1.87 $-$ 2.24  &                         & 0.596 $-$ 1.23 & $<$ 0.14 & \\
NGC~5408      & PL & $\Gamma =$ 2.39 & 4.20 $\times$ 10$^{-4}$ & 0 & & 261 / 269 \\
              &    &  2.33 $-$ 2.48  &                         & $<$ 0.29 & & \\
              & RS & kT $=$ 2.02     & 1.64 $\times$ 10$^{-3}$ & 0 & 0.03 & 343 / 268 \\
              &    &  1.86 $-$ 2.19  &                         & $<$ 0.04 & $<$ 0.12 & \\
\\
\tableline
\tablebreak
\multicolumn{3}{l}{Elliptical Galaxies}
\\
NGC~4374 & PL & $\Gamma =$ 3.29\tablenotemark{g} & 1.13 $\times$ 10$^{-3}$ & 1.69\tablenotemark{g} & & 447 / 216 \\
         &    &                                  &                         &                       & & \\
         & RS & kT $=$ 0.79  & 4.94 $\times$ 10$^{-3}$ & 1.10 & 0.08 & 372 / 215 \\
         &    &  0.75 $-$ 0.82 &                       & 0.71 $-$ 1.59 & 0.06 $-$ 0.11 & \\
NGC~4406 & PL & $\Gamma =$ 5.72\tablenotemark{g} & 9.42 $\times$ 10$^{-3}$ & 5.15\tablenotemark{g} & & 1189 / 386 \\
         &    &                                  &                         &                       & & \\
         & RS & kT $=$ 0.80  & 6.69 $\times$ 10$^{-3}$ & 1.41 & 0.40 & 459 / 385 \\
         &    & 0.77 $-$ 0.82 &                        & 0.86 $-$ 2.10 & 0.31 $-$ 0.57 & \\
NGC~4552 & PL & $\Gamma =$ 2.24 & 2.70 $\times$ 10$^{-4}$ & 0.558 & & 404 / 248 \\
         &    &  2.06 $-$ 2.44  &                         & 0.088 $-$ 1.07 & & \\
         & RS & kT $=$ 0.76     & 2.11 $\times$ 10$^{-3}$ & 2.23  & 0.10 & 454 / 247 \\
         &    &  0.71 $-$ 0.81  &                         & 1.45 $-$ 3.35 & 0.07 $-$ 0.16 & \\  
\enddata
\vfill\eject
 
\tablenotetext{}{NOTE: Ranges for parameters are given for 90\% confidence 
   for one interesting parameter ($\Delta\chi^2 =$ 2.7).}
\tablenotetext{a}{Emission model used in the spectral fit:
   PL $=$ power-law model and RS $=$ Raymond-Smith thermal plasma model.}
\tablenotetext{b}{Photon spectral index $\Gamma$ for the power-law 
   model and gas temperature kT for the RS model.}
\tablenotetext{c}{Normalization for the PL and RS models.  Units for
   the PL model are photons~keV$^{-1}$~cm$^{-2}$~s$^{-1}$ at 1 keV.  For the
   RS model, the normalization is equal to 10$^{-14}$ / (4$\pi$D$^2$) 
   $\int$ n$_e$ n$_H$ dV, where D is the distance to the X-ray source in cm,
   n$_e$ (n$_H$) is the electron (hydrogen) density in units of cm$^{-3}$, 
   and V is the volume of the X-ray emitting source.}
\tablenotetext{d}{Effective neutral hydrogen absorbing column density.}
\tablenotetext{e}{Abundance of metals in the thermal plasma, relative to
  cosmic abundances.}
\tablenotetext{f}{$\chi^2$ value for the fit and number of degrees of
  freedom (no. spectral bins - no. of fit parameters).}
\tablenotetext{g}{Fit is too poor to use $\chi^2$ statistics to find
  error range (i.e., $\chi^2_\nu >$ 2.0 ).}
\end{deluxetable}



\addtocounter{table}{2}
\addtocounter{page}{3}



\clearpage
\begin{deluxetable}{lcccc}
\scriptsize
\tablecaption{Inner Radius of Accretion Disk for Spiral Galaxies
  \label{tabrincostheta}}
\tablewidth{0pt}
\tablehead{
\colhead{Galaxy} & 
\colhead{kT$_{in}$\tablenotemark{a}} &
\colhead{R$_{in}$ $(\cos\theta)^{{1}\over{2}}$\tablenotemark{a}} & 
\colhead{M(R$_{in}(\cos\theta)^{{1}\over{2}} =$ 3r$_g$)\tablenotemark{b}} & 
\colhead{M(BH)\tablenotemark{c}} \\
\colhead{Name} & 
\colhead{(keV)} &
\colhead{(km)} &
\colhead{(M$_\odot$)} &
\colhead{(M$_\odot$)} \\
} 
\startdata
M33 \#1      & 1.15 & 37.1 (29.7 $-$ 49.1) & 4.2 & 116.7 \\
M33 \#2      & 1.15 & 40.8 (34.4 $-$ 50.6) & 4.6 & 117.6 \\
NGC~1313 \#2 & 1.34 & 33.4 ( $<$ 75.8 )    & 3.8 & 590 \\
NGC~5408     & 0.13 & 18090 (2191 $-$ 62559) & 2044 & $\sim$8$-$9 $\times$ 10$^3$ \\
\enddata
 
\tablenotetext{a}{Temperature kT$_{in}$ at the inner radius of the accretion
  disk (column 1) and the inner radius of accretion disk times the 
  square root of the cosine
  of the inclination angle (column 2), as given by the normalization
  of the multicolor disk blackbody model.  Ranges are given based on 90\%
  confidence ranges of the normalizations to the DBB+PL fits in Table 6}
\tablenotetext{b}{Mass of the central black hole if the inner radius
  r$_{in}(\cos\theta)^{{1}\over{2}}$ is assumed to be at 3 Schwarzschild
  radii (the innermost stable circular orbit for a Schwarzschild black
  hole).  Note that this mass is assumed incorrect (see text section
  6.3.2)}
\tablenotetext{c}{Mass of the central black hole, as given by the normalization of
  the bulk motion model.}
\end{deluxetable}


\clearpage
\begin{deluxetable}{lccccccccrrr}
\scriptsize
\tablecaption{Bulk Motion Fits to Spiral Galaxy ASCA Spectra
  \label{tabascabm} }
\tablewidth{0pt}
\tablehead{
\colhead{Galaxy} & 
\colhead{kT\tablenotemark{a}} &
\colhead{$\Gamma$\tablenotemark{b}} &
\colhead{$\log$ f\tablenotemark{c}} &
\colhead{Normalization\tablenotemark{d}} &
\colhead{N$_H$\tablenotemark{e}} &
\colhead{$\chi^2$ / d.o.f.\tablenotemark{f}} \\
\colhead{Name} & 
\colhead{(keV)} &
\colhead{} &
\colhead{} &
\colhead{} &
\colhead{(10$^{21}$ cm$^{-2}$)} &
\colhead{} \\
} 
\startdata
M33 \# 1 & 0.46 & 1.96 & -1.78 $\times$ 10$^{-5}$ & 1.58 $\times$ 10$^{-4}$ & 0 & 687 / 593 \\
         & 0.45 $-$ 0.47 & 1.87 $-$ 2.06 &        &                         & $<$ 0.026   & \\
M33 \# 2 & 0.47 & 2.00 & -1 $\times$ 10$^{-5}$    & 1.75 $\times$ 10$^{-4}$ & 0 & 873 / 674 \\
         & 0.46 $-$ 0.48 & 1.93 $-$ 2.09  &       &                         & $<$ 0.016   & \\
NGC~1313 \#1 & 0.12 & 1.80 & -1.61 $\times$ 10$^{-5}$ & 2.17 $\times$ 10$^{-5}$ & 1.39 &  238 / 257 \\
         & 0.11 $-$ 0.15 & 1.72 $-$ 1.86 &      &                              & 0.94 $-$ 1.89   & \\
NGC~1313 \#2 & 0.33 & 2.17 & -3.06 $\times$ 10$^{-5}$ & 2.09 $\times$ 10$^{-5}$ & 0 & 194 / 196 \\
         & 0.29 $-$ 0.35 & 2.03 $-$ 2.33 &            &                         & $<$ 0.726 & \\
NGC~5408 & 0.12 & 2.35 & -2.42 $\times$ 10$^{-5}$ & 2.03 $\times$ 10$^{-5}$ & 0.60 & 256 / 267 \\
         & 0.11 $-$ 0.14 & 2.22 $-$ 2.45 &        &                         & 0.28 $-$ 0.95 & \\
\enddata
 
\tablenotetext{}{NOTE: Ranges for parameters are given for 90\% confidence 
   for one interesting parameter ($\Delta\chi^2 =$ 2.7).}
\tablenotetext{a}{Temperature of the disk for BM model.}
\tablenotetext{b}{Photon index for power-law model.}
\tablenotetext{c}{Logarithm of the fraction $f$ of the disk blackbody emission
   which is scattered for the BM model.}
\tablenotetext{d}{Normalization of the BM model.}
\tablenotetext{e}{Effective neutral hydrogen absorbing column density.}
\tablenotetext{f}{$\chi^2$ value for the fit and number of degrees of
  freedom (no. spectral bins - no. of fit parameters).}
\end{deluxetable}

\clearpage

 
\begin{deluxetable}{llcccccl}
\scriptsize
\tablecaption{Additional Notes on Sample of Nearby Galaxies
  \label{tabclass} }
\tablewidth{0pt}
\tablehead{
\colhead{Galaxy} & \colhead{Other} & \colhead{X-ray} &  
\colhead{Class\tablenotemark{b}} & \colhead{Ref.\tablenotemark{b}} & 
\colhead{M$_\bullet$\tablenotemark{c}} & \colhead{Ref.\tablenotemark{c}} & 
\colhead{Notes\tablenotemark{d}} \\
\colhead{Name} & \colhead{Name} & \colhead{Source?\tablenotemark{a}} & 
\colhead{} & \colhead{} &
\colhead{log M$_\odot$} & \colhead{} &
\colhead{} \\
} 
\startdata
NGC~205  & M110 & no  & ...  &    & ... &    &             \\
NGC~221  & M32  & yes & ...  &    & 6.5 & 1,2 &                  \\
NGC~300  &      & no  & ...  &    & ... &    &                  \\
NGC~598  & M33  & yes & H~II & 1  & ... &    &                  \\
NGC~1313 &      & yes & H~II & 2  & ... &    &                 \\
NGC~1396  &     & no  & ...  &    & ... &    &      \\
AM~0337-353 &   & no  & ...  &    & ... &     &  \\
IC~342     &    & yes & H~II & 1  & ... &    &   \\
NGC~1705  &     & no  & H~II & 3  & ... &    & SB ref 1, NED H~II \\
NGC~2403  &     & yes & H~II & 1  & ... &     &  \\
Ho~II &         & yes & ...  &    & ... &     & normal ref 2 \\
ESO~495-G21 & He~2-10 &yes & ... &   & ... &      & PN-like galaxy NED notes \\
UGC~5086 &      & no & ... &      & ... &     &  dwarf \\
NGC~3031  & M81 & yes & S1.5,LINER &  1,2 & ... &    &  \\
Ho~IX &         & no & ... &     & ... &    &  dwarf  \\
Leo~A &         & no & ... &     & ... &    &  dwarf \\  
NGC~3990 &      & no & ... &     & ... &    &   \\
NGC~4138  &     & no & S1.9 & 1  & ... &    &  \\
NGC~4150     &  & yes & ... &    & ... &     &  \\
NGC~4190    &   & yes & ... &    & ... &     &   \\
UGCA~276 & DDO~113 & no & ... &  & ... &     &   \\
UGC~7356 &      &  no & ... &    & ... &     &   \\
NGC~4278 &      & yes & L1.9,Sy1 & 1,3 & 9.2 & 2  &   \\
NGC~4286 &      & no & ... &     & ... &     &   \\
\tableline
\tablebreak
NGC~4374 & M84  & yes & L2  & 1 & 9.2 & 3 &  \\
IC~3303  &      &  no & ... &   & ... &    &  \\
NGC~4387  &     &  no & ... &   & ... &    &  \\
NGC~4406  & M86 & yes & ... &   & ... &    &  \\
NGC~4449   &    & yes & H~II & 1,2  & ... &     & SB ref 2 \\
NGC~4485  &     & yes & H~II & 1  & ... &   & \\
IC~3540   &     & no  & ... &   & ... &    &  \\
NGC~4552 &  M89 & yes & T2: & 1 & 8.7 &  2  &  \\
NGC~4569 &  M90 & yes  & T2,LINER & 1,2 & ... &    &   \\
NGC~4627 &      & no  & ... &     & ... &    & \\
NGC~5204 &      & yes & H~II & 1 & ... &    & \\
NGC~5408 &      & yes & ... &   & ... &    &  \\
NGC~6822 &      & yes & ... &   & ... &    &  \\
NGC~6946 &      & yes & H~II & 1 & ... &    &  \\
NGC~7320 &      & no & ... &    & ... &    & \\ 
\enddata

 
\tablenotetext{a}{Was a compact X-ray source found in the current
paper?}
\tablenotetext{b}{Nuclear optical spectrum classification.
References: (1) Ho, Filippenko \& Sargent 1995;
(2) Kinney et al. 1993; (3) NED
}
\tablenotetext{c}{Central Dark Mass as determined by dynamical means.
References: (1) Bender et al. 1996; (2) Magorrian et al. 1998; (3) Bower
at al. 1998}
\tablenotetext{d}{Notes on galaxies.  References: (1) Lamb et al. 1985; 
(2) Kinney et al. 1993}
\end{deluxetable}

%
%
%


\clearpage
\begin{deluxetable}{lccccccccrrr}
\scriptsize
\tablecaption{Unabsorbed X-ray Luminosities of Central X-ray Sources
  \label{tabascalum} }
\tablewidth{0pt}
\tablehead{
\colhead{Galaxy} & 
\colhead{F(2$-$10 keV)\tablenotemark{a}} &
\colhead{L$_X$(2$-$10 keV)\tablenotemark{b}} \\
\colhead{Name} & 
\colhead{(erg~s$^{-1}$~cm$^{-2}$)} &
\colhead{(erg~s$^{-1}$)} \\
} 
\startdata
M33 \# 1     & 1.02 $\times$ 10$^{-11}$ & 5.98 $\times$ 10$^{38}$ \\
M33 \# 2     & 1.11 $\times$ 10$^{-11}$ & 6.51 $\times$ 10$^{38}$ \\
NGC~1313 \#1 & 1.86 $\times$ 10$^{-12}$ & 3.05 $\times$ 10$^{39}$ \\
NGC~1313 \#2 & 1.06 $\times$ 10$^{-12}$ & 1.74 $\times$ 10$^{39}$ \\
NGC~5408     & 6.36 $\times$ 10$^{-13}$ & 1.83 $\times$ 10$^{39}$ \\
\tableline
NGC~4374     & 6.86 $\times$ 10$^{-13}$ & 2.32 $\times$ 10$^{40}$ \\
NGC~4406     & 5.42 $\times$ 10$^{-13}$ & 1.83 $\times$ 10$^{40}$ \\
NGC~4552     & 5.73 $\times$ 10$^{-13}$ & 1.94 $\times$ 10$^{40}$ \\
\enddata
 
\tablenotetext{a}{Unabsorbed 2$-$10 keV flux for: total model (spiral 
  galaxies M33, NGC~1313 and NGC~5408) or hard component (elliptical
  galaxies (NGC~4374, NGC~4406 and NGC~4552).}
\tablenotetext{b}{X-ray luminosities for the fluxes given in column 2
   and the distances given in Table 3?}.

\end{deluxetable}
\clearpage

\clearpage

\figcaption{Histogram of Galaxy Type for galaxies with detected X-ray
sources (upper panel) and galaxies in entire sample (lower panel).  
For clarity, the galaxy ESO~495-G21 (type 90.0) has been
omitted from the plot. \label{fig1}}

\figcaption{Histogram of the 0.2$-$2.4 keV X-ray luminosity of the
compact X-ray sources.  Galaxy type for each galaxy is indicated on the
plot.  \label{fig2}}

\figcaption{Histogram of the separation between the X-ray position of
the detected X-ray sources and the photometric center of the galaxy, in
units of {\bf (a)} arcseconds and {\bf (b)} parsecs.  \label{fig3}}

\figcaption{Plot of logarithm of the 0.2$-$2.4 keV X-ray luminosity and
the angular separation of the X-ray source from the photometric center.
Those galaxies with previous evidence of nuclear activity are labelled
as a filled rectangle, while unclassified galaxies are labelled with an
open rectangle. \label{fig4}}

\figcaption{Plot of the logarithm of the ROSAT X-ray luminosity and the
blue luminosity (from Tables 5 and 1, respectively).  Elliptical
galaxies are labelled as filled rectangles, spiral galaxies are labelled
as filled triangles, and dwarf and irregular galaxies are labelled as
open triangles.  The solid line marks the predicted X-ray luminosity
of X-ray binaries (from Canizares et al. 1987) for early-type galaxies.
\label{fig5}}

\end{document}